\documentclass[usenatbib,usegraphicx]{mn2e}

\def\apj{ApJ}
\def\apjl{ApJL}
\def\apjs{ApJS}
\def\aap{A\&A}
\def\aapr{A\&A Review}
\def\mnras{MNRAS}
\def\aj{AJ}
\def\nat{Nature}
\def\araa{ARA\&A}
\def\prd{Ph. Rv. D}
\def\apss{ApSS}
\def\jgr{Journal of Geophysical Research}

\def\cm{\textrm{cm}}

\def\erg{\textrm{erg}}
\def\kpc{\textrm{kpc}}
\def\pc{\textrm{pc}}
\def\Mpc{\textrm{Mpc}}
\def\Gpc{\textrm{Gpc}}
\def\Kelv{\textrm{K}}

\def\ergps{\textrm{ergs}~\textrm{s}^{-1}}

\def\kms{\textrm{km}~\textrm{s}^{-1}}
\def\gcm2{\textrm{g}~\textrm{cm}^{-2}}
\def\ergscm3{\textrm{ergs}~\textrm{s}^{-1}~\textrm{cm}^{-3}}
\def\ergcm3{\textrm{ergs}~\textrm{cm}^{-3}}
\def\gscm2{\textrm{g}~\textrm{s}^{-1}~\textrm{cm}^{-2}}
\def\ergcmK34{\textrm{erg}~\textrm{cm}^{-3}~\textrm{K}^{-4}}
\def\cc{\textrm{cm}^{-3}}
\def\cms31{\textrm{cm}^{-3}~\textrm{s}^{-1}}
\def\cmg21{\textrm{cm}^{2}~\textrm{g}^{-1}}

\def\phcm2s1{\textrm{photons}~\textrm{cm}^{-2}~\textrm{s}^{-1}}

\def\JUnits{\textrm{ergs}~\textrm{cm}^{-2}~\textrm{s}^{-1}~\textrm{sr}^{-1}}
\def\eV{\textrm{eV}}

\def\TeV{\textrm{TeV}}
\def\Hz{\textrm{Hz}}
\def\kHz{\textrm{kHz}}
\def\MHz{\textrm{MHz}}
\def\GHz{\textrm{GHz}}

\def\sec{\textrm{s}}
\def\yr{\textrm{yr}}

\def\Gyr{\textrm{Gyr}}

\def\muGauss{\mu\textrm{G}}

\def\GeVs1cm3{\textrm{GeV}~\textrm{s}^{-1}~\textrm{cm}^{3}}
\def\log{\textrm{log}}
\def\2phn{\phn\phn}
\def\3phn{\phn\phn\phn}
\def\4phn{\phn\phn\phn\phn}
\def\12phn{\4phn\4phn\4phn}
\def\Msun{\textrm{M}_{\sun}}
\def\Lsun{\textrm{L}_{\sun}}
\newcommand{\mean}[1]{\ensuremath{\langle #1 \rangle}}

\title[Limits on Extragalactic Sub-MHz Radio]{The End of the Rainbow: What Can We Say About the Extragalactic Sub-Megahertz Radio Sky?}
\author[Lacki]{Brian C. Lacki$^{1,2}$\thanks{E-mail:lacki@astronomy.ohio-state.edu}\\
$^1$Department of Astronomy, The Ohio State University, 140 W. 18th Avenue, Columbus, OH 43210, USA, lacki@astronomy.ohio-state.edu\\
$^2$Center for Cosmology \& Astro-Particle Physics, The Ohio State University, 191 W. Woodruff Ave., Columbus, OH 43210, USA}

\begin{document}

\maketitle

\begin{abstract}
The Galactic disc is opaque to radio waves from extragalactic sources with frequencies $\nu$ less than $\sim 3\ \MHz$.  However, radio waves with kHz, Hz, and even lower frequencies may propagate through the intergalactic medium (IGM).  I argue that the presence of these waves can be inferred by using the Universe as our detector.  I discuss possible sub-MHz sources and set new non-trivial upper limits on the energy density of sub-MHz radio waves in galaxy clusters and the average cosmic background.  Limits based on five effects are considered: (1) changes in the expansion of the Universe from the radiation energy density (2) heating of the IGM by free-free absorption; (3) radiation pressure squeezing of IGM clouds by external radio waves; (4) synchrotron heating of electrons in clusters; and (5) Inverse Compton upscattering of sub-MHz radio photons.  Any sub-MHz background must have an energy density much smaller than the CMB at frequencies below 1 MHz.  The free-free absorption bounds from the Lyman-$\alpha$ forest are potentially the strongest, but are highly dependent on the properties of sub-MHz radio scattering in the IGM.  I estimate an upper limit of $6 \times 10^4\ \Lsun\ \Mpc^{-3}$ for the emissivity within Lyman-$\alpha$ forest clouds in the frequency range $5 - 200$ Hz.  The sub-MHz energy density in the Coma cluster is constrained to be less than $\sim 10^{-15}\ \ergcm3$.  At present, none of the limits is strong enough to rule out a maximal $T_b = 10^{12}\ \Kelv$ sub-MHz synchrotron background, but other sources may be constrained with a better knowledge of sub-MHz radio propagation in the IGM.
\end{abstract}

\begin{keywords}
radio continuum: general -- radiation mechanisms: non-thermal -- intergalactic medium -- miscellaneous
\end{keywords}

\section{Introduction}
Whenever a new wavelength window has been opened on the electromagnetic spectrum, it has led to new discoveries \citep{Harwit81,Lawrence07}.  In the past century, almost the entire electromagnetic spectrum has been explored, with few gaps between $10~\MHz$ and $100~\TeV$.  The cosmic electromagnetic backgrounds at most of these frequencies\footnote{Although the extragalactic Extreme Ultraviolet (EUV; $100 - 912$ \AA) radiation is easily absorbed by neutral hydrogen, EUV has been detected from AGNs and galaxy clusters at wavelengths of $\la 160\ {\rm \AA}$ (see the EUV review by \citealt*{Bowyer00}).  Closer to the Lyman limit, not even a Galactic background has been detected yet \citep{Edelstein01}, and the Galactic neutral hydrogen would much more effectively block incoming EUV radiation.  However, the extragalactic ionizing background at a variety of redshifts is indirectly measured by the ionization state of Lyman-$\alpha$ forest clouds (e.g., \citealt*{Bajtlik88}; \citealt{Shull99}).} have now been measured \citep[as reviewed by][]{Ressell90,Fukugita04,Trimble06}.  Searches are underway for very-high energy photons, including those with PeV \citep{Chantell97,Borione98,Schatz03} and even EeV energies \citep[e.g.,][]{Auger08}.  These searches face several challenges.  At PeV energies, extragalactic searches are hampered by $\gamma\gamma$ opacity, in which PeV photons interact with the CMB to produce $e^+e^-$ pairs, and the Universe is highly opaque at these energies \citep*{Moskalenko06}.  EeV photons are predicted to exist (e.g., \citealt*{Wdowczyk72,Gelmini08}), and the Universe is possibly transparent to several Mpc at these energies \citep{Protheroe96}, but very low number statistics are a problem.  Finally, above $10^{24} \eV$, the Galaxy becomes completely opaque, as single photons pair produce $e^+e^-$ off the Galactic magnetic field \citep{Stecker03}.  The only other frontier in terms of photon energy is at the other end of the electromagnetic spectrum, the lowest frequency radio waves.

The Universe is filled with tenuous plasma, the intergalactic medium (IGM), which prevents radio propagation below the plasma frequency $\nu_P = \sqrt{n_e e^2 / (\pi m_e)}$.  Waves below this frequency evanesce within one wavelength, reflecting off the medium.  The mean baryonic density of the Universe is $\mean{n_b} \approx 2.5 \times 10^{-7} (1 + z)^3~\cm^{-3}$, although regions in the IGM may have higher or lower densities.  For IGM with electron density $\delta_e \mean{n_b}$, the plasma frequency is
\begin{equation}
\label{eqn:nuP}
\nu_P = 4.5 \delta_e^{1/2} (1 + z)^{3/2}~\Hz.
\end{equation}
The density distribution in the Universe can be approximated as a lognormal distribution, with most of the volume being relatively empty \citep{Coles91}.  Roughly $90\%$ of the Universe's volume at $z = 0$ is predicted to have $\delta_e \ga 0.002$ \citep{Bi97}, which corresponds to a plasma frequency of $\nu_{P} = 0.2~\Hz$.  This is essentially the low end of the cosmic electromagnetic spectrum, below which no electromagnetic wave can ever travel.\footnote{There is a loophole, since electromagnetic waves below the plasma frequency will evanesce with a scale of about one wavelength: if the electromagnetic wave has a cosmologically long wavelength, it can stretch over the entire Universe.  Hawking radiation from the accelerating expansion of the Universe actually does have cosmological wavelengths \citep[e.g.,][]{Gibbons77}, but the energy density of Hawking radiation is negligible.}

Although there might be electromagnetic radiation with $\nu \la \Hz$ in the Universe, direct observations at these frequencies are impossible.  Observations down to $10\ \MHz$ will soon be routine with LOFAR \citep{Rottgering03}.  However, the Earth's ionosphere has a typical plasma frequency of $10\ \MHz$, depending on the time of day and other conditions.  Similarly, the Moon may have an ionosphere with a plasma frequency of a few hundred kHz, which sets a lower limit to the frequency of lunar-based observatories \citep{Jester09}.  Satellites have measured the Galactic radio emission down to $\sim 100\ \kHz$ \citep{Brown73,Novaco78}.  Space-based observatories near the Earth, such as the proposed ALFA \citep{Jones00a,Jones00b}, can potentially observe down to $30\ \kHz$, which is the plasma frequency of the Solar wind at Earth's orbit.  Further out from the Sun, the plasma frequency continues to drop, and the \emph{Voyager} probes took advantage of this to detect kHz emission \citep{Kurth84}.  However, the interstellar medium itself has a plasma frequency of 2 kHz, which serves as a hard limit for direct observations of Galactic and extragalactic sources.

The situation is even worse for extragalactic and distant galactic sources.  The warm ionized medium (WIM) forms a disc with scale height $h \approx 1~\kpc$ and a typical density of $n_{\rm WIM} \approx 0.01~\cm$.  The WIM is opaque to low frequency radio emission because of free-free absorption.  Although this absorption has uses for tomography of the Galactic ISM \citep{Peterson02}, it prevents all direct extragalactic observations at $\nu \la 3\ \MHz$.  There may be a few low density `chimneys' that allow in some lower frequency radio emission \citep{Jester09}, but for practical purposes, the extragalactic sky at the lowest radio frequencies will be shrouded from our direct view for the foreseeable future.

But there are ways around this limit to locate or place bounds on \emph{extragalactic} sub-MHz radio sources.  Free-free absorption and other opacity sources become more effective at low frequency; this is why they prevent sub-MHz radiation from reaching Earth.  Yet, this also means that the lowest frequency radio waves are tightly coupled with the matter in the Universe.  Observations of intergalactic matter therefore constrain these radio waves.  Synchrotron absorption and Inverse Compton scattering also place bounds on sub-MHz radio in regions with cosmic rays and magnetic fields, like galaxy clusters.  I will argue that sub-MHz radio emission does not need to reach us, because \emph{the Universe is our detector}.

I will first discuss our expectations for the sub-MHz sky (\S~\ref{sec:SubMHzSky}), including postulated sub-MHz radio sources (\S~\ref{sec:SubMHzSources}), the IGM phases that can interact with sub-MHz radio waves (\S~\ref{sec:IGMPhases}), and sub-MHz radio propagation through the IGM (\S~\ref{sec:IGMPropagation}).  I then set new limits on the extragalactic background at sub-MHz frequencies.  The first limit I consider is the weak bound from the expansion history of the Universe (\S~\ref{sec:OmegaR}).  The second bound is from the heating of the IGM by free-free absorption of sub-MHz radio waves (\S~\ref{sec:FFAbsorption}).  A third bound on incident radiation at the lowest frequencies comes from the radiation force exerted on IGM clouds (\S~\ref{sec:CloudCrush}).  Two more bounds can be set for clusters with cosmic rays: bounds on synchrotron heating by low frequency radio waves (\S~\ref{sec:SynchHeat}) and bounds on Inverse Compton upscattered radio waves (\S~\ref{sec:ICBound}).  Finally, if extragalactic photons with energy $\ga 10^{20} \eV$ are ever detected, they will set extremely strong limits on the kHz to MHz radio background (\S~\ref{sec:UHELimits}).

My goal throughout this paper is to derive upper bounds on the sub-MHz emission with order of magnitude accuracy, where possible with current knowledge.  In some cases even this is not possible -- the radio scattering properties of the IGM over large distances are not well known, and this is a huge source of uncertainty in arguments that rely on radiative transfer.  Future theoretical work may reduce these uncertainties.  Throughout this work, I consider the time-averaged sub-MHz background at $z \approx 0$, though these methods may be applied to other redshifts.

\section{The Extragalactic Sub-MHz Sky}
\label{sec:SubMHzSky}

\subsection{Sources}
\label{sec:SubMHzSources}
\subsubsection{Synchrotron emission}
The MHz to GHz radio emission of galaxies and some galaxy clusters is dominated by synchrotron emission from cosmic ray (CR) electrons and positrons (\citealt{Condon92} reviews the synchrotron radio emission of non-AGN galaxies, while \citealt{Ferrari08} reviews the synchrotron radio emission of galaxy clusters).  This emission should continue down to lower frequencies, where mostly lower energy electrons and positrons are radiating.  Non-relativistic electrons and positrons emit at the cyclotron frequency
\begin{equation}
\label{eqn:nuB}
\nu_{B} = \frac{e B}{2 \pi m_e c} = 2.8~\Hz \left(\frac{B}{\muGauss}\right)
\end{equation}
At GHz wavelengths, the radio spectrum falls off steeply as $\nu^{-\alpha}$, where $\alpha \approx 0.7$ for galaxies and $1 \le \alpha \le 1.5$ for clusters, implying greater spectral densities at low frequency and greater power at low frequency for clusters.  

However, the radio spectra of galaxies are unlikely to continue down to the sub-MHz range for a number of reasons.  First, CR spectra tend to be flattened at low energies, because of escape, ionization, or bremsstrahlung losses, or simply because the low energy CR electrons are not old enough to be cooled by synchrotron emission.  Second, free-free absorption will prevent sub-MHz radiation from escaping galaxies, just as it prevents it from entering the Galaxy.  Third, at the Razin frequency, the index of refraction of the plasma the CRs traverse suppresses the relativistic beaming of synchrotron emission.  Below the Razin frequency, the synchrotron spectrum falls off precipitously.  In the Galaxy, the Razin frequency is $\sim 0.1~\MHz$.  \citet{Protheroe96} calculated the kHz to GHz radio background, including synchrotron emission.  In terms of energy density, the spectrum falls off at frequencies below about a MHz.  

The prospects for strong sub-MHz synchrotron radiation from clusters are a bit more encouraging.  Galaxy clusters have much lower densities than galaxies, so bremsstrahlung and ionization losses are weaker.  The low density and high hot temperature of the gas that fills clusters also means that free-free absorption will be much less effective at blocking sub-MHz radio emission (discussed further in \S~\ref{sec:IGMPropagation}).  Finally, the low density means that the Razin frequency will be much lower.  For typical galaxy cluster densities and magnetic field strengths, the Razin cutoff is \citep{Schlickeiser02}
\begin{equation}
\label{eqn:nuRazin}
\nu_R = 18.5~\kHz\ n_{e,-3} B_{\rm \mu G}^{-1}.
\end{equation}
with $n_{e,-3} = n_e / (10^{-3} \cc)$ and $B_{\rm \mu G} = B / (\mu G)$.  We may expect that synchrotron emission in clusters may extend all the way down to kHz frequencies.

In addition, there can be low frequency synchrotron emission from more diffuse structures.  There appear to be radio structures associated with poor galaxy groups or filaments of galaxies \citep{Delain06,Brown09}.  Shocks from large scale structure formation are also predicted to produce low frequency radio emission (\citealt{Waxman00}; \citealt*{Keshet04}).

Synchrotron emission is expected to have a brightness temperature limited to $T_b \la 10^{11 - 12}~\Kelv$, because of Inverse Compton scattering by the synchrotron emitting particles among other concerns \citep[e.g.,][]{Kellermann69,Readhead94}.  This would imply a low energy density of synchrotron radio emission at low frequencies,
\begin{equation}
\nu u_{\nu,\rm max synch} \la 1.3 \times 10^{-25}\ \ergcm3\ \nu_{\rm kHz}^3 T_{b,12},
\end{equation}
where $\nu_{\rm kHz} = \nu / {\rm kHz}$, and $T_{b,12} = T_b / 10^{12} \Kelv$.  This limit only applies to steady-state sources; transients can have much greater brightness temperatures.

\subsubsection{Processes in Plasmas}
A number of plasma processes generate low frequency radio emission.  These processes are often coherent and can evade the $T_b = 10^{12}~\Kelv$ limit on incoherent synchrotron emission.  Many of these processes are observed in the Sun \citep[see the review by][]{Dulk85} and the Solar System.  Langmuir waves in plasma can be converted into radio emission near the plasma frequency and its first harmonic \citep[e.g.,][]{Dulk85}.  \emph{Voyager} has detected 1 -- 4 kHz radio emission \citep{Kurth84}, now known to be emitted somewhere near the heliopause \citep{Kurth03}.  Since its frequency is comparable to the plasma frequency of the ISM and the Solar Wind at the heliopause, it is generally believed to be powered by Langmuir wave conversion (e.g., \citealt{Macek96}; \citealt*{Treumann98}).  Coherent plasma processes have been suggested to generate radio emission in AGNs, but it is not clear that coherent emission can escape from the AGNs \citep{Melrose99}.

When the cyclotron frequency (eq.~\ref{eqn:nuB}) exceeds the plasma frequency (eq.~\ref{eqn:nuP}), and when the electron velocity distribution function grows with perpendicular velocity, the cyclotron maser instability can also generate bright radio emission \citep[e.g.,][]{Winglee86,Treumann06}.  The cyclotron maser mechanism is believed to be the main source of Jupiter's bright decametric radio emission \citep*{Hewitt81} and Earth's auroral kilometric radiation \citep{Wu79,Zarka98}, and may operate around exoplanets \citep*[e.g.,][]{Bastian00}.  The cyclotron maser is also believed to operate in stars, being responsible for the Sun's microwave spike bursts and stellar radio bursts \citep{Melrose82}, and coherent periodic radio emission from brown dwarfs, though at higher frequencies \citep{Hallinan07,Hallinan08}.

The same cosmic rays that produce the bulk of the synchrotron emission can also produce radiation by interacting with plasma.  Electrons travelling through an inhomogeneous plasma can emit transition radiation below the Razin frequency, caused by a varying index of refraction \citep{Fleishman92,Fleishman95}.  Transition radiation is expected to be the main source of Galactic emission below $\sim 100~\kHz$ \citep{Fleishman95}.

\subsubsection{Exotic possibilities}
\label{sec:ExoticSources}
\emph{Pulsar waves} -- Pulsars are expected to generate magnetic dipole radiation of enormous amplitude as they spin down \citep{Gunn69}.  The pulsar wave would have a frequency equal to the pulsar spin frequency, which would be less than a kHz for a millisecond pulsar (MSP) and $\sim 30\ \Hz$ for the Crab Pulsar.  However, the environment around pulsars is expected to be filled with dense plasma, which might prevent the escape of these pulsar wave into intergalactic space \citep{Goldreich69}.  

There are roughly 40000 MSPs in the Galactic disc (\citealt{Ferrario07}; \citealt*{Story07}; \citealt{Lorimer08}), and 1000 in globular clusters \citep{Heinke05}.  Assuming they have an average luminosity of $10^{34}\ \ergps$, each lasting 10 Gyr \citep{Lorimer08}, and scaling from the stellar mass of the Milky Way ($\sim 5 \times 10^{10}\ \Msun$; \citealt{Flynn06}) to total stellar mass of the local Universe ($\sim 5 \times 10^8\ \Msun\ \Mpc^{-3}$; \citealt{Cole01,Hopkins06}), MSPs can provide a maximum energy density of $4 \times 10^{-20}\ \ergcm3$ from galactic discs and $1 \times 10^{-21}\ \ergcm3$ from globular clusters.  Young pulsars are born at approximately the supernova rate, or $\Gamma_{SN} = 1 / (30\ \yr)$ for Milky Way-like galaxies, and have a spin energy of $5 \times 10^{49}~\erg$, yielding a maximum $\sim 30\ \Hz$ background of $6 \times 10^{-18}\ \ergcm3$.  The young pulsars and MSPs in the galactic discs are unlikely to contribute to the extragalactic background, however, because of free-free absorption in the galactic discs and the interstellar plasma frequency is often above 1 kHz.  Some pulsars may contribute if they are kicked out of their galactic disc during the supernova.  Most likely, intergalactic absorption will reduce the energy densities further from these maximum estimates, so these energy densities are upper limits.

\emph{Conversion of gravitational waves} -- Possibly one of the most exciting sources of low frequency radio waves would be gravitational waves (GWs).  Typical sources of gravitational waves have frequencies of less than about a kilohertz.  \citet*{Marklund00} showed that if they passed through a magnetized plasma, they can generate radio waves of the same frequency.  Even if non-linear plasma effects generate radio waves of higher frequency, the resulting frequency can still be below MHz \citep{Marklund00}.  Gravitational waves can also excite cyclotron resonances \citep{Papadopoulos02}.  Black hole ringdown is predicted to be an important source of these converted gravitational waves \citep{Clarkson04}.  

LIGO's frequency range ($\sim 100 - 1000\ \Hz$) is above the plasma frequency of several IGM phases, meaning that sources detectable by LIGO may excite radio waves.  These sources can include compact object mergers and cosmic strings.  Transients like merging stellar-mass black holes will produce bursts of gravitational waves.  The cosmic energy density in gravitational waves is roughly $u = \Gamma t_{\rm burst} u_{\rm max}$, where $\Gamma$ is the rate of observed gravity wave bursts, $t_{\rm burst} \approx \nu^{-1}$ is the duration of the burst, and $u_{\rm max}$ is the gravitational wave energy density within the burst itself.  This can be evaluated approximately as $u \approx c^2 \nu^2 h^2 \Gamma / G$, where $h$ is the strain at Earth.  LIGO has constrained the rate of $h \ga 10^{-21} \Hz^{-1/2}$ bursts with $64\ \Hz \la \nu \la 2000\ \Hz$ to be $\Gamma_{\rm bright} < 3.6\ \yr^{-1}$ \citep{Abbott09a}, implying that the cosmic energy density of bright gravitational wave bursts is $u_{\rm bright} \la 4 \times 10^{-16}\ \ergcm3 \nu_{500}^2 h_{-21}^2$, where $\nu_{500} = \nu / (500\ \Hz)$ and $h_{-21} = h / (10^{-21}\ \Hz^{-1/2})$.  An associated radio background is likely to have a much smaller energy density, because only a small fraction of the gravitational waves is converted into radio and because intergalactic absorption will convert the radio background into heat.  However, faint gravitational wave bursts are not as well constrained by LIGO.

LIGO has also constrained the $100\ \Hz$ stochastic gravitational wave background (SGWB) to have $u < 5.7 \times 10^{-14}\ \ergcm3$, assuming they have a power law spectrum, which should apply to cosmic strings \citep{Abbott09b}.  Again, any related radio background would probably have a much lower energy density, especially considering intergalactic absorption.

Gravitational waves detectable by LISA, such as those from supermassive black hole mergers, will be below the plasma frequency of any reasonable density medium.  They will be detectable only if they excite radio waves of much higher frequency.

\subsection{The Intergalactic Medium: Our Detector for Sub-MHz Radio Waves}
\label{sec:IGMPhases}
The phases of the intergalactic medium (IGM) are believed to fall into three main categories: a low density ($\delta \la 10$), relatively cold ($\sim 10^4\ \Kelv$) photoionized phase visible as the Lyman $\alpha$ forest; a denser, ($\delta \approx 5 - 100$) collisionally ionized phase that is hotter ($T \approx 10^5 - 10^7\ \Kelv$), known as the Warm-Hot Intergalactic Medium (WHIM); and condensed structures ($\delta \ga 100$) associated with galaxies and clusters \citep{Dave99}.  Since each of these phases have different temperatures and densities, they interact with radio waves in different ways, with different absorption properties and thermal contents.  

I shall consider the effects that a sub-MHz background would have on these IGM phases.  A large enough background would alter the properties of these phases, and would indirectly be in conflict with observation.  I now discuss the phases I will be considering in further detail (properties summarised in Table \ref{table:IGMSummary}):
\begin{itemize}
\item \emph{The Lyman-$\alpha$ forest}: The Lyman-$\alpha$ forest is detected as weak Lyman-$\alpha$ absorption lines (by definition $N({\rm H I}) < 10^{17.2} \cm^{-2}$, but typically $N({\rm H I}) < 10^{14.5} \cm^{-2}$) in the spectra of bright background objects, such as quasars.  The Lyman-$\alpha$ forest is particularly prominent at $z \ga 2$, where it dominates the baryonic content of the Universe.  However, similar clouds exist at $z = 0$, where they make up $\sim 30\%$ of the baryonic gas (e.g., \citealt{Dobrzycki02}; \citealt*{Penton04}).  These clouds represent a `fluctuating Gunn-Peterson trough' -- that is, they are denser regions in the background IGM, which by itself is too rarefied to appear as absorption at low $z$ \citep{Bi97}.  Lyman-$\alpha$ clouds are somewhat clustered near large-scale structure; most do not appear to be associated with individual galaxies, and some are in voids (\citealt*{Penton02}; \citealt{Stocke07}).  I consider Lyman-$\alpha$ forest clouds with overdensities $\delta = 0.1$ and $1.0$, which are typical of Lyman-$\alpha$ forest clouds\footnote{Note that while Lyman $\alpha$ clouds are `overdense' with respect to the background IGM that fills most of the \emph{volume} of the Universe, they can have less than the average baryonic density of the Universe ($\delta < 1$).  This is because the baryonic density in most of the volume is much below the average baryonic density.}, with temperatures of $10^4 \Kelv$ \citep{Bi97}, and with radii of 100 kpc.

\item \emph{Warm-Hot Intergalactic Medium}: The WHIM is one of the more elusive phases of the IGM, despite being presumed to contain about $\approx 30\%$ of the baryonic content of the Universe at $z = 0$ (see the review by \citealt{Bregman07}; also see \citealt{Cen99}).  Its presence is none the less inferred from UV and X-ray absorption lines of highly ionized species, such as collisionally ionized O VI absorbers (e.g., \citealt*{Tripp00}; \citealt{Cen01,Danforth06,Danforth08}).  The WHIM is believed to be associated with structure formation; as gas gravitationally collapses to form large-scale structure, it is shocked and heats up \citep{Dave01}.  It should be clustered near large-scale structure but generally would not be associated with individual galaxies.  I consider WHIM clouds with overdensities $\delta = 5$ and $100$, and with a temperature of $10^6\ \Kelv$, which are typically expected values.  I assume WHIM clouds have radii of 1 Mpc.

\item \emph{Cool WHIM}: Numerical simulations by \citet{Kang05} predict the existence of a `cool WHIM', which is collisionally ionized and with the same densities as the WHIM, but colder ($10^3 \Kelv \la T \la 10^5 \Kelv$).  The existence of O VI absorption systems apparently associated with cold Ly$\alpha$ absorbers may be a sign that this phase actually exists \citep{Tripp08}.  I consider a cool WHIM cloud to have an overdensity $\delta = 10$, a temperature of $10^4 \Kelv$, and a radius of 1 Mpc.

\item \emph{Cluster gas}: An extremely hot ($10^8 \Kelv$) phase of gas pervades galaxy clusters, and is easily visible in free-free X-ray emission.  I consider galaxy clusters with gas densities of $10^{-4} \cc$ ($2 \times 10^{12}\ \Msun\ \Mpc^{-3}$ or $\delta = 400$) and $10^{-3} \cc$ ($2 \times 10^{13}\ \Msun\ \Mpc^{-3}$ or $\delta = 4000$) \citep*{Mohr99}.  The latter density is more typical of rich clusters.  Cluster gas temperature is assumed to be $10^8~\Kelv$ \citep{David93}, and the cluster gas has a radius of 1 Mpc in this paper.

\item \emph{Metal-line absorbers}: Relatively dense photoionized gas is also present in the inner tens of kiloparsecs around $L^{\star}$ galaxies, manifesting itself as strong Ly$\alpha$ and metal absorption lines in quasar spectra.  I consider strong Mg II absorbers with $n = 0.01\ \cc$ and $R = 1~\kpc$ with $T = 10^4 \Kelv$ \citep{Ding03,Ellison04}.  Weak metal line absorbers are also present around smaller, dwarf galaxies \citep{Stocke04,Keeney06}, and possibly in intergalactic space.  Photoionized weak metal-line absorbers include dense structures with $n \approx 0.01\ \cc$ and $R \approx 10\ \pc$, detected in low ionization lines like Mg II and Fe II (\citealt*{Rigby02}; \citealt{Richter09}); kiloparsec-scale structures with $n \approx 10^{-3} \cc$ that are observed in C IV at $z \approx 1$ and Mg II at $z \approx 0$ (\citealt*{Rauch01}; \citealt{Charlton03,Milutinovic06,Narayanan05}); and possibly high-ionization absorbers with $n \approx 10^{-5} \cc$ and $R \approx 100~\kpc$ seen in high ionization lines, such as C IV haloes at $z \approx 0$ (\citealt*{Chen01}; \citealt{Lacki10}) and intergalactic O VI clouds at $z \approx 1$ \citep{Zonak04,Tripp08,Oppenheimer09}.  I consider examples of all these structures, with a typical photoionized temperature of $T = 10^4 K$.

\end{itemize}

These phases span a wide range of density, temperature, and locations near structures; from Lyman $\alpha$ clouds that often appear in voids, to metal line absorbers that are concentrated very near galaxies.  However, most of the volume of the Universe is filled with voids with extremely low density gas ($\delta \ll 1$).  Thus, if there was sub-MHz radio that somehow filled \emph{only} voids, then it may not have a great effect on detected IGM phases.  Even though there are some Lyman-$\alpha$ clouds in voids, they would have a plasma frequency greater than the background IGM; thus they would reflect incident radio waves of the lowest frequencies rather than being directly heated by them.  Even so, the pressure exerted by the exterior radiation background on the cloud could still have an effect if it was too high.

\begin{table*}
\begin{minipage}{170mm}
\label{table:IGMSummary}
\caption{IGM Phase Summary.}
\begin{tabular}{lccccccccc}
\hline
Phase & $R$ & $n_e [\delta_e]$ & $\nu_P$ & $T$ & $\Lambda_N^a$ & $t_{\rm therm}$ & $t_{\rm dyn}$ & $t_{\rm sound}$ & $t_{\rm IGM}$ \\ 
& ($\Mpc$) & ($\cm^{-3}$) & ($\Hz$) & ($\Kelv$) & ($\erg~\cm^3$) & ($\Gyr$) & ($\Gyr$) & ($\Gyr$) & ($\Gyr$)\\
\hline
Background IGM$^b$ & $1000$ & $5.0 \times 10^{-10} [0.002]$ & $0.2$ & $10^4$ & ... & ... & $4000$ & $6 \times 10^4$ & $10$\\
Underdense Ly$\alpha$ forest & $0.1$ & $2.5 \times 10^{-8} [0.1]$ & $1.4$ & $10^4$ & $-2.3 \times 10^{-23}$ & $200$ & $600$ & $6$ & $6$\\
Ly$\alpha$ forest & $0.1$ & $2.5 \times 10^{-7} [1]$ & $4.5$ & $10^4$ & $-1.7 \times 10^{-23}$ & $30$ & $200$ & $6$ & $6$\\
Low density WHIM & $1$  & $1.25 \times 10^{-6} [5]$ & $10$ & $\sim 10^6$ & $1.4 \times 10^{-23}$ & $800$ & $80$ & $6$ & $6$\\
High density WHIM & $1$  & $2.5 \times 10^{-5} [100]$ & $45$ & $\sim 10^6$ & $2.1 \times 10^{-23}$ & $30$ & $20$ & $6$ & $6$\\
Cool WHIM$^b$ & $1$ & $10^{-6} [4]$ & $9.0$ & $10^4$ & $-1.5 \times 10^{-23}$ & $9$ & $90$ & $60$ & $9$\\
Low density cluster & $1$ & $10^{-4} [400]$ & $90$ & $10^8$ & $3.7 \times 10^{-23}$ & $400$ & $9$ & $0.6$ & $0.6$\\
High density cluster & $1$ & $10^{-3} [4000]$ & $280$ & $10^8$ & $3.0 \times 10^{-23}$ & $40$ & $3$ & $0.6$ & $0.6$\\
Strong Mg II absorber  & $0.001$ & $10^{-2} [40000]$ & $900$ & $10^4$ & $-6.5 \times 10^{-25}$ & $0.02$ & $1$ & $0.06$ & $0.02$\\
Dense weak Mg II absorber & $10^{-5}$ & $10^{-2} [40000]$ & $900$ & $10^4$ & $-6.5 \times 10^{-25}$ & $0.02$ & $1$ & $6 \times 10^{-4}$ & $6 \times 10^{-4}$\\
Weak Mg II/C IV absorber & $0.001$ & $10^{-3} [4000]$ & $280$ & $10^4$ & $-1.2 \times 10^{-24}$ & $0.07$ & $3$ & $0.06$ & $0.06$\\
C IV/O VI Galaxy Halo$^b$ & $0.1$ & $10^{-5} [40]$ & $28$ & $10^4$ & $-1.1 \times 10^{-23}$ & $1$ & $30$ & $6$ & $1$\\
\hline
\end{tabular}
\\$^a$  Cooling rates for intergalactic clouds at $z = 0$ for $Z = 0.1 Z_{\sun}$ metallicity from \citet{Wiersma09}.  Negative cooling rates imply net heating.\\
$^b$  These phases have not been confirmed observationally.
\end{minipage}
\end{table*}

The sensitivity reached by an IGM cloud-detector depends on its effective integration time, or the time $t_{\rm IGM}$ that an IGM cloud-detector would be expected to last.  There are four relevant time-scales for most IGM phases\footnote{There are additional possible time scales in some phases.  For cluster gas, there is also the Alfven crossing time, $t_{\rm Alfven} = R / v_A$ where $v_A = B / \sqrt{4 \pi \rho}$ is the Alfven speed.  The Alfven crossing time is $t_{\rm Alfven} = 14\ \Gyr R_{\rm Mpc} B_{\rm \mu G}^{-1} n_{-3}^{1/2}$, which is longer than the sound crossing time for clusters.}:
\begin{enumerate}
\item \emph{Dynamical time-scale} -- The time for the cloud to collapse under self-gravity.  I assume a dynamical time of $t_{\rm dyn} \approx 1 / \sqrt{G \rho}$, or
\begin{equation}
t_{\rm dyn} \approx 190\ \Gyr\ \delta_e^{-1/2}
\end{equation}
\item \emph{Thermal time-scale} -- The IGM can cool radiatively, but is also being photoionized by the extragalactic UV background.  We can define a thermal time-scale
\begin{equation}
t_{\rm therm} = \frac{3 k T}{n_e |\Lambda_N|}
\end{equation}
In the IGM, the cooling coefficient $\Lambda_N$ can be negative because of the photoionizing background.  I have used the values of $\Lambda_N$ from \citet*{Wiersma09}.
\item \emph{Sound crossing time-scale} -- In general, this is \emph{not} equal to the the dynamical time-scale, since IGM clouds like those in the Lyman-$\alpha$ forest are not virialized.  The sound speed is $c_s = \sqrt{10 k T / (3 m_H)}$, giving a typical sound crossing time of
\begin{equation}
\label{eqn:tSound}
t_{\rm sound} \approx 59\ \Gyr\ R_{\rm Mpc} T_4^{-1/2},
\end{equation}
where $R_{\rm Mpc}$ is the radius of the cloud in Mpc, and $T_4 = T / 10^4\ \Kelv$. 
\item \emph{Age of the Universe} -- For simplicity, I use $t_H \approx 10\ \Gyr$.
\end{enumerate}
I choose the shortest of these times to be $t_{\rm IGM}$, the expected survival time of an IGM cloud.  These times are compared in Table~\ref{table:IGMSummary}.

Finally, there is the \emph{relativistic} phase of the IGM, the cosmic rays (CRs).  Cosmic ray electrons (and positrons) are known to exist in galaxy clusters, through their synchrotron radio emission and the Inverse Compton emission in X-rays \citep{Ferrari08,Rephaeli08}.  Presumably there also is a more diffuse cosmic ray background that pervades the Universe.  A large fraction of cosmic rays, both electrons (and positrons) and protons, escape from smaller, less-dense galaxies \citep{Aublin06}.  Dark matter annihilation or decay may also inject CRs into the IGM \citep[see][]{Chen04}.  However, no truly diffuse extragalactic CR background has ever been detected, except at the highest energies.

\subsection{Propagation of Sub-MHz Radio through the IGM}
\label{sec:IGMPropagation}

\begin{figure}
\centerline{\includegraphics[width=8cm]{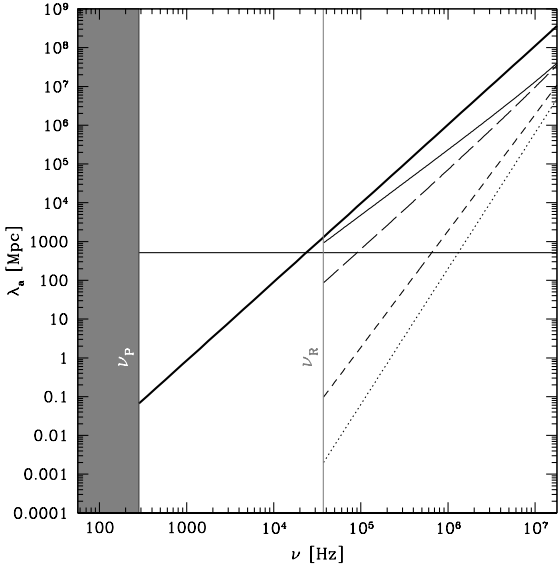}}
\caption{Absorption path lengths for the Coma Cluster at low frequencies, with $B = 0.5\ \muGauss$ and $n = 10^{-3} \cc$.  The thick black line is free-free absorption.  The thin diagonal lines are for synchrotron absorption; solid assumes the spectrum falls precipitously at low energies ($\nu_C \la 30~\MHz$), long dashed assumes a flat spectrum at low energies, short dashed assumes an $E^{-2}$ spectrum continuing down to $\gamma = 1$, and dotted assumes an $E^{-3}$ spectrum continuing down to $\gamma = 1$.  The horizontal line is Thomson scattering.  Synchrotron absorption dominates at frequencies above the Razin cutoff ($\nu_R$), but free-free absorption remains below that frequency.  The cluster is optically thick at $\nu \la \kHz$ from free-free absorption, and near the Razin cutoff from synchrotron absorption if the low energy CR electron spectrum is steep. \label{fig:ClusterAbsorption}}
\end{figure}

Sub-MHz radio waves suffer a number of absorption and scattering processes as they traverse the IGM.

\emph{Thomson scattering} -- Since the IGM is ionized, it contains free electrons that will Thomson scatter incident electromagnetic radiation of all radio frequencies.  However, the optical depth to Thomson scattering is very low, since the IGM is so rarefied:
\begin{equation}
\lambda_s = 1950\ \delta_e^{-1}\ \Gpc.
\end{equation}
Therefore, Thomson scattering is unlikely to be of any significance in the IGM.

\emph{Free-free absorption} -- Free-free absorption becomes effective at lower frequencies, and would blanket out the low frequency Universe.  The free-free absorption coefficient is
\begin{equation}
\label{eqn:AlphaNuFF}
\alpha_{\nu} = 0.018 T^{-3/2} \overline{g_{\rm ff}} n_e^2 \nu^{-2} \cm^5~\Hz^2~\Kelv^{3/2}.
\end{equation}
The Gaunt factor $\overline{g_{\rm ff}}$ for radio frequencies above the plasma frequency and below $kT/h$ is \citep{Scheuer60,Spitzer78}
\begin{equation}
\overline{g_{\rm ff}} = 9.77 [1 + 0.195 \log_{10} T_4 - 0.130 \log_{10} \nu_{\rm MHz}].
\end{equation}
Over the frequencies considered in this paper, $\overline{g_{\rm ff}} \approx 10 - 20$.

The absorption mean free path of sub-MHz radio in the IGM is roughly
\begin{equation}
\lambda_a = 19\ \Mpc\ T_4^{3/2} \overline{g_{\rm ff,15}}^{-1} \delta_e^{-2} \nu_{\rm kHz}^2.
\end{equation}
The background IGM phase I consider ($\delta_e = 0.002$) is transparent across Gpc distances down to $\sim 15\ \Hz$.  However, a radio wave could run into denser structures that are more opaque and be absorbed.  Typical Lyman-$\alpha$ and WHIM clouds become transparent above about $10 - 1000~\Hz$, assuming scattering within the clouds is negligible.  The strong Mg II absorbers are opaque for $\nu \la 250\ \kHz$, even without scattering.  Thus, the time $\lambda_a / c$ that a radio photon can travel is limited by the time until it reaches one of these denser structures.  Somewhat perversely, scattering can actually increase the absorption time, by trapping photons in IGM phases with low absorption.

\emph{Synchrotron absorption} -- Synchrotron absorption would be effective in galaxy clusters, where a high content of CRs and magnetic fields are present.  The synchrotron absorption coefficient is
\begin{equation}
\alpha_{\nu} = -\frac{c^2}{8 \pi \nu^2} \int dE P_s(\nu,E) E^2 \frac{\partial}{\partial E} \left[\frac{N(E)}{E^2}\right],
\end{equation}
where $P_s (\nu,E)$ is the synchrotron power at frequency $\nu$ (above the Razin cutoff; equation~\ref{eqn:nuRazin}) of a single electron/positron at energy $E$:
\begin{equation}
P_s(\nu,E) = \frac{\sqrt{3} e^3 B \sin \alpha}{m c^2} \frac{\nu}{\nu_c} \int_{\nu/\nu_c}^{\infty} K_{5/3} (\xi) d\xi
\end{equation}
with
\begin{equation}
\nu_C = \frac{3 \gamma^2 e B \sin \alpha}{4 \pi m c}.
\end{equation}

The synchrotron absorption depends on the CR electron spectrum in clusters.  It is possible to estimate the electron spectrum for $10^3 \la \gamma \la 10^4$ from the MHz to GHz synchrotron radio they emit using the monochromatic approximation \citep{Felten66,Schlickeiser02}:
\begin{equation}
N(E) = 9 \frac{e \sin \alpha}{m^2 \sigma_T c^4 B} \gamma^{-1} L_{\nu}.
\end{equation}
The disadvantage of this method is that there is a minimum CR energy probed by synchrotron radio emission, because there is a minimum radio frequency observed $\nu_{\rm minobs}$:
\begin{equation}
\gamma_{\rm minobs} = 1930 \left(\frac{B}{\muGauss}\right)^{-1/2} \left(\frac{\nu_{\rm minobs}}{10~\MHz}\right)^{1/2}
\end{equation}
We can, of course, use models for the spectrum at lower energy, but the presence of sub-MHz radio would actually alter the low energy spectrum through synchrotron heating (as described in \S~\ref{sec:SynchHeat}).  When possible, we should use clusters with fluxes detected at the lowest possible frequencies.  The Coma cluster radio halo has observations at frequencies ranging from 30 MHz to 5 GHz, so I have used it as an example \citep*{Thierbach03}.  Other clusters are typically observed at GHz frequencies, though more clusters will be seen at lower frequencies (down to $\sim 15\ \MHz$) with LOFAR. 

In Figure~\ref{fig:ClusterAbsorption}, I show the importance of Thomson scattering (\emph{grey}), free-free absorption (\emph{red}), and synchrotron absorption (\emph{blue}) for the Coma Cluster at low frequencies.  The synchrotron absorption depends sensitively on the low energy ($\gamma \la 1000$) CR electron/positron spectrum.  I show that if there are no electrons below $\gamma_{\rm minobs}$ (\emph{solid}), synchrotron absorption is relatively weak at lower frequencies.  The steeper the spectrum below $\gamma_{\rm minobs}$, the greater the synchrotron absorption, as can be seen if the spectrum has a constant $dN/dE$ for $\gamma \le \gamma_{\rm minobs}$ (\emph{dotted}) and if the spectrum goes as $dN/dE \propto E^{-2}$ or $E^{-3}$ for $\gamma \le \gamma_{\rm minobs}$ (\emph{dashed}).

\emph{Plasma inhomogeneity scattering} -- Finally, sub-MHz radio may scatter off plasma inhomogeneities in the IGM \citep{Cohen74}.  Suppose the plasma inhomogeneities have a spectrum $P_n (\vec{q}) = C_n^2 q^{-\alpha}$, where the wavenumber $q = 2\pi / \ell$ ranges from a minimum $q_0$ (large wavelength) to a maximum $q_1$ (small wavelength).  The power $P_n$ is defined by $\int P_n (\vec{q}) d^3 \vec{q} = \mean{\Delta n_e^2}$.  For an isotropic Kolmogorov spectrum of inhomogeneities ($\alpha = 11/3$), where $\ell_0 \gg \ell_1$, the coefficient $C_n^2$ is
\begin{equation}
C_n^2 \approx \frac{2^{1/3}}{3 \pi^{4/3}} \frac{\mean{\Delta n_e^2}}{\ell_0^{2/3}}.
\end{equation}

The image of a point source is smeared over an angle $\theta_c = c / (2 \pi \nu \ell_c)$, where
\begin{equation}
\label{eqn:rhoCScattering}
\ell_c \approx \left[4 \pi^2 \frac{e^2}{m \nu^2} s C_n^2 f(\alpha)\right]^{-1/(\alpha-2)}
\end{equation}
where $s$ is the path length through the medium, and
\begin{equation}
f(\alpha) = 2^{2 - \alpha} \frac{\Gamma(\alpha/2) (\alpha - 2)}{\Gamma(2 - (\alpha/2))},
\end{equation}
but equation~\ref{eqn:rhoCScattering} holds only if $2 < \alpha < 4$ and $\ell_1 \ll \ell_c \ll \ell_0$ (\citealt{Lee75}; \citealt*{Cordes85}).  If we take the scattering mean free path $\lambda_s$ to be the distance when the image of a point source is smeared out over $\theta_c = 1$, then $\ell_c = c / (\pi^2 \nu)$.  For a Kolmogorov spectrum of turbulence ($\alpha = 11/3$), 
\begin{eqnarray}
\label{eqn:mfpScattering}
\lambda_s & \approx & 1.5 \times 10^{-5}\ \Mpc\ \ell_{0,\rm Mpc}^{2/3} \Delta_n^{-1} n_{e,-3}^{-2} \nu_{\rm MHz}^{11/3}\\
 & \approx & 240\ \Mpc\ \ell_{0,\rm Mpc}^{2/3} \Delta_n^{-1} \delta_e^{-2} \nu_{\rm MHz}^{11/3},
\end{eqnarray}
where $\Delta_n = \mean{\Delta n_e^2} / n_e^2$.  The scattering coefficient is then just defined to be $\delta_{\nu} = \lambda_s^{-1}$.  

There is a great deal of uncertainty in the derived path length.  Most importantly, equation~\ref{eqn:mfpScattering} holds \emph{only if} $\ell_1 \ll c / (\pi^2 \nu) \ll \ell_0$.  In the Milky Way, $\ell_1 \approx 10^7 \cm$ \citep{Spangler90}; if this is true in the IGM, then these formulae are valid only if $\nu \ll 300~\Hz$.   The two natural scales for $\ell_1$ are the ion Larmor radius and the ion inertial length \citep{Spangler90}, both of which are expected to be very large in the IGM.  The mean free path also depends on the outer scale of the largest inhomogeneities $\ell_0$ and their amplitude $\mean{\Delta n_e^2}$.  In our Galaxy, the inhomogeneities continue merely up to $\sim \pc$ scales rather than the size of the Galaxy \citep{Haverkorn08}.  On the other hand, the amplitude of the fluctuations is almost certainly less than the mean ISM density in our Galaxy.   

The presence of scattering in the IGM would trap sub-MHz radio near their sources of emission.  The scattering optical depth of a medium with radius $R = R_{\rm Mpc} \Mpc$ is $\tau_s = R / \lambda_s$, or, if the estimate in equation~\ref{eqn:mfpScattering} holds,
\begin{eqnarray}
\tau_s & \approx & 6.5 \times 10^4 R_{\rm Mpc} \ell_{0,\rm Mpc}^{-2/3} \Delta_n n_{e,-3}^{2} \nu_{\rm MHz}^{-11/3}  \\
 & \approx & 4.1 \times 10^8 R_{\rm Mpc} \ell_{0,\rm Mpc}^{-2/3} \Delta_n \delta_e^{2} \nu_{\rm kHz}^{-11/3}.
\end{eqnarray}
In an optically thick medium, the sub-MHz radio must diffuse with a typical time scale $t_{\rm diff} = 3 R^2 / (c \lambda_s)$, or
\begin{eqnarray}
\label{eqn:tDiff}
t_{\rm diff} & \approx & 6.4 \times 10^{11}\ \yr\ R_{\rm Mpc}^2 \ell_{0,\rm Mpc}^{-2/3} \Delta_n n_{e,-3}^{2} \nu_{\rm MHz}^{-11/3}  \\
 & \approx & 4.0 \times 10^{15}\ \yr\ R_{\rm Mpc}^2 \ell_{0,\rm Mpc}^{-2/3} \Delta_n \delta_e^{2} \nu_{\rm kHz}^{-11/3}.
\end{eqnarray}

Thus, based on the short distance found in equation~\ref{eqn:mfpScattering} and long diffusion times, even if we could somehow place a sub-MHz detector outside of the Galactic WIM, we still may not see most of the sub-MHz radio sources in the Universe.  However, since this is not an issue at present, the scattering can serve to our advantage.  Sub-MHz radio is trapped near its source; therefore, if we observe the effects of sub-MHz radio (as described in following sections), we know that its source is nearby.  Trapping \emph{localises} sub-MHz radio for us, and we do not simply have to consider a diffuse background.

It is possible to define an effective absorption mean free path, which is the displacement a radio photon can travel before being absorbed.  When scattering is strong, the effective absorption mean free path is much shorter than $\lambda_a$, because the scattering traps the photon near its source.  The effective absorption mean free path is $\lambda_{\rm a,eff} = 1 / \sqrt{\alpha_{\nu} (\alpha_{\nu} + \delta_{\nu})}$.  If equation~\ref{eqn:mfpScattering} holds and free-free absorption is the main absorption source, then $\delta_{\nu} \gg \alpha_{\nu}$, and
\begin{equation}
\label{eqn:lambdaAbsEff}
\lambda_{\rm a,eff} = 6.9 \times 10^4\ \Mpc\ \ell_{0,\rm Mpc}^{1/3} \Delta_n^{-1/2} \delta_e^{-2} \overline{g_{\rm ff,15}}^{-1/2} T_4^{3/4} \nu_{\rm MHz}^{17/6}.
\end{equation}
We see that $\lambda_{\rm a,eff}$ steeply falls as the frequency decreases.  

\section{Limit from $\Omega_R$}
\label{sec:OmegaR}
Since the extragalactic ultra-low frequency radio background cannot be detected from the Solar System, it can be thought of as dark radiation, a relativistic component of the Universe not directly seen.  Like other postulated kinds of dark radiation, sub-MHz radio must have a gravitational effect on the expansion of the Universe.  So cosmological bounds on the radiation energy density of the Universe also limit extragalactic sub-MHz radio.

Most cosmological limits are very stringent for the early Universe ($z \ga 1000$), but do not rule out the growth of $\Omega_R$ late in the Universe.  Since presumably sub-MHz radio would be radiated by non-thermal processes after the CMB has decoupled and structure has begun forming, limits from the CMB and Big Bang Nucleosynthesis are largely irrelevant.  Type Ia supernovae data give us the limit $\Omega_R \la 0.2$ for $z \la 1000$, or $u \la 1.7 \times 10^{-9}\ \erg\ \cm^{-3}$ at $z = 0$ \citep{Zentner02}.  This argument applies to the globally averaged sub-MHz radio background, and not to specific, compact regions like clusters.  

Unlike other more exotic forms of dark radiation, sub-MHz radio does interact with intergalactic matter.  These interactions are then visible from Earth, and are generally much more constraining.

\section{IGM Heating from Free-Free Absorption}
\label{sec:FFAbsorption}

Just as free-free absorption prevents sub-MHz radio from entering the Galaxy, it should also be present in intergalactic space.  The free-free absorption converts low frequency radio waves into IGM heat, which can affect the dynamical \citep{Ciotti04} or thermal state of the gas.  Therefore, the IGM serves as a low-frequency radio detector: if the radio background was too high, the IGM would heat up to a higher temperature than observed.  Low frequency radio emission would create \emph{anomalous heating} in the IGM; the lack of such heating puts an upper limit on the radio background.

\subsection{Energy density limits}
\label{sec:FFUBound}
Consider a small sphere of IGM plasma immersed in an isotropic bath of sub-MHz radio with steady intensity $J_{\nu}$ for a time $t_{\rm IGM}$ (\S~\ref{sec:IGMPhases}), during which we expect the IGM not to be otherwise heated or cooled significantly.  The heating rate of the plasma parcel is
\begin{equation}
\label{eqn:IGMHeating}
\dot{Q} \approx \pi A \ell \int \frac{-dJ_{\nu}}{d\ell} d\nu,
\end{equation}
where $A$ is the surface area of the sphere, $\ell$ is the mean path length through the sphere, and $\frac{dJ_{\nu}}{d\ell}$ is the mean change in radio intensity per path length.  If the brightness temperature of the radio bath is much greater than the temperature of the plasma, then $\frac{dJ_{\nu}}{d\ell} \approx -\alpha_{\nu} J_{\nu}$, given an absorption coefficient $\alpha_{\nu}$.  We shall suppose that we are considering a very small parcel, so that there is no scattering and $\ell \approx R$, the radius of the sphere; then $A \ell \approx 3 V$, where $V$ is the volume of the sphere.  To avoid excess heating, 
\begin{equation}
\label{eqn:ExcessHeatParcel}
\dot{Q} \le \frac{3 NkT_{\rm IGM}}{2 t_{\rm IGM}}.
\end{equation}
In this equation, $N$ is the total number of particles and is approximately equal to twice $N_e$, the number of electrons.

Equation~\ref{eqn:IGMHeating} applies to a small parcel of gas, and the relevant $J_{\nu}$ is the intensity \emph{within that parcel}.  However, the intensity of the radio bath may vary drastically throughout the cloud if it is externally illuminated and optically thick to free-free absorption, in which case the radio intensity is very low deep inside the cloud. 

\emph{Spatially homogeneous $J_{\nu}$ case} --  First, consider the case when the radio background is constant throughout an IGM cloud.  This assumption is appropriate if either the cloud is optically thin (to both scattering and absorption), or if internal objects distributed nearly continuously\footnote{Continuous means here that the separation between sources is less than the effective absorption mean free path within the cloud.} within the cloud are the source of the radio bath, as would be the case for synchrotron or transition emission emitted by CRs in the cloud.

Suppose the radio bath consists of a nearly constant bump at frequency $\nu$ and with width $\Delta \nu \la \nu$.  Then conditions~\ref{eqn:IGMHeating} and~\ref{eqn:ExcessHeatParcel} gives us
\begin{equation}
\label{eqn:JFFLimitInt}
\Delta\nu J_{\nu} \la \frac{nkT_{\rm IGM}}{2 \pi t_{\rm IGM} \alpha_{\nu}}.
\end{equation}
Assuming that $n_e = \delta_e \mean{n_b} = n / 2$, the free-free absorption coefficient gives the following upper bounds:
\begin{eqnarray}
\label{eqn:FFBumpLimitJ}
\Delta\nu J_{\nu} & \la & 2.1 \times 10^{-11} \frac{T_4^{5/2} \nu_{\rm kHz}^2}{t_{10} \overline{g_{\rm ff,15}} \delta_e}~\JUnits\\
\label{eqn:FFBumpLimitu}
\Delta\nu u_{\nu} & \la & 8.7 \times 10^{-21} \frac{T_4^{5/2} \nu_{\rm kHz}^2}{t_{10} \overline{g_{\rm ff,15}} \delta_e}~\erg~\cm^{-3},
\end{eqnarray}
with $T_4 = T / (10^4~\Kelv)$, $\nu_{\rm kHz} = \nu / (\kHz)$, $\overline{g_{\rm ff,15}} = \overline{g_{\rm ff}} / 15$, and $t_{10} = t_{\rm IGM} / (10~\Gyr)$.

In Table~\ref{table:FFNuUNuLimits}, I consider various phases of the IGM and set limits on the homogeneous radio background within these phases.  The best limits come from dense, cold phases of the IGM, such as galactic metal-line absorbers.  Hot phases like the WHIM and cluster gas place relatively poor limits on the radio background.  The corresponding brightness temperatures of the allowed radio backgrounds are high enough that ignoring the thermal emission of the IGM cloud is appropriate.

Although I assumed the spectral intensity was a bump of width $\Delta\nu \la \nu$ when calculating the limits in equations~\ref{eqn:FFBumpLimitJ} and~\ref{eqn:FFBumpLimitu}, the heating can be integrated over any desired spectrum in equation~\ref{eqn:IGMHeating}.  In particular, for a power law of intensity $J_{\nu} = J_0 (\nu / \nu_0)^q$ in the frequency range $\nu_{\rm min} \ll \nu \ll \nu_{\rm max}$,
\begin{equation}
\label{eqn:FFPowerLawLimitJ}
J_0 \la  2.1 \times 10^{-17} \frac{T_4^{5/2}}{\delta_e t_{10} \overline{g_{\rm ff,15}}(\nu_0)} \times \left\{ \begin{array}{ll}
		\displaystyle (q - 1.1) \frac{\nu_0^{q - 0.1}}{\nu_{\rm max}^{q - 1.1}} & (q > 1.1)\\
	        \displaystyle (1.1 - q) \frac{\nu_0^{q - 0.1}}{\nu_{\rm min}^{q - 1.1}} & (q < 1.1)\\
                \displaystyle \nu_0 \left({\rm ln}\frac{\nu_{\rm max}}{\nu_{\rm min}}\right)^{-1} & (q = 1.1),
		\end{array} \right.
\end{equation}
where $J_0$ is in cgs units ($\JUnits$) and I have used the approximation that the Gaunt factor goes as $\nu^{-0.1}$ with a normalisation $\overline{g_{\rm ff,15}} (\nu_0) = g_{\rm ff} (\nu_0) / 15$ at $\nu_0$.  

\emph{Externally illuminated, optically thick case} -- Now suppose an optically thick cloud has no internal sub-MHz luminosity, but is immersed in an \emph{external} radio bath.  Since the cloud is optically thick, the interior of the cloud will at first be unaffected by the radio bath.  The exterior of the cloud will heat up through free-free absorption.  As it heats up, it will become more transparent to free-free absorption (equation~\ref{eqn:AlphaNuFF}), both because of its higher temperature and because it can expand away and dilute.  The next layer in is then exposed to the external radio bath, so that the cloud is effectively eroded by the external radio bath.  

I assume that the intensity of the radio bath is homogeneous and equal to its external value $J_{\nu}$ within one absorption effective mean free path $\lambda_{\rm a,eff} = [\alpha_{\nu,\rm ff} (\alpha_{\nu,\rm ff} + \delta_{\nu})]^{-1/2}$ of the cloud exterior, where $\delta_{\nu} = \lambda_s^{-1}$ is the scattering coefficient.  Then the time for this layer to evaporate is $t_{\rm evap} (\lambda_{\rm a,eff}) = Q / \dot{Q}$, or from equation~\ref{eqn:ExcessHeatParcel},
\begin{equation}
t_{\rm evap} (\lambda_{\rm a,eff}) \approx \frac{3 n_e k T}{\pi \alpha_{\nu,\rm ff} (\Delta\nu J_{\nu})}
\end{equation}
assuming the spectrum of the radio bath is a narrow bump of width $\Delta\nu$.

The time for the entire cloud to evaporate is therefore $t_{\rm evap} (R) \approx (R / \lambda_{\rm a,eff}) t_{\rm evap} (\lambda_{\rm a,eff})$, or
\begin{equation}
t_{\rm evap} (R) \approx \frac{3 R n_e k T}{\pi \Delta\nu J_{\nu}} \sqrt{1 + \frac{\delta_{\nu}}{\alpha_{\nu,\rm ff}}}.
\end{equation}
Note that there is now no frequency dependence if there is no scattering ($\delta_{\nu} = 0$).  Since the cloud is optically thick, it absorbs all of the incident radiation, which is converted into heat.  Scattering slows down the evaporation process, since the incident radiation cannot penetrate as deeply into the cloud at any given time.  

In order for the cloud to survive in the radio bath, we require that $t_{\rm evap} (R) \ga t_{\rm IGM}$.  For typical values in the IGM, the limits on external radio backgrounds for optically thick clouds are\footnote{Note that the bounds on $\Delta\nu J_{\nu}$ and $\Delta\nu u_{\nu}$ in the externally illuminated, optically thick case (eq.~\ref{eqn:FFBumpLimitJThick} and~\ref{eqn:FFBumpLimituThick}) do not match the homogeneous case (eq.~\ref{eqn:FFBumpLimitJ} and~\ref{eqn:FFBumpLimitu}) when $\tau_{\rm a,eff} = 1$ (at either $\nu_{\rm cal} (R)$ given in eq.~\ref{eqn:nuCalR}, or at $\nu_{\rm cal,s} (R)$ given in eq.~\ref{eqn:nuCalRs}).  The discrepancy is caused by a geometrical factor: in the homogeneous case, the heated volume is a sphere with radius $R$, but the optically thick case assumes the heated volume is a thin shell with radius $R$ and depth $\lambda_{\rm a,eff}$, which will overestimate the volume if $\lambda_{\rm a,eff} = R$.}
\begin{eqnarray}
\label{eqn:FFBumpLimitJThick}
\Delta\nu J_{\nu} & \la & 3.2 \times 10^{-12} \frac{\delta_e T_4 R_{\rm Mpc}}{t_{10}} \sqrt{1 + \frac{\delta_{\nu}}{\alpha_{\nu,\rm ff}}}\\
\label{eqn:FFBumpLimituThick}
\Delta\nu u_{\nu} & \la & 1.4 \times 10^{-21} \frac{\delta_e T_4 R_{\rm Mpc}}{t_{10}} \sqrt{1 + \frac{\delta_{\nu}}{\alpha_{\nu,\rm ff}}},
\end{eqnarray}
where $\Delta\nu J_{\nu}$ and $\Delta\nu u_{\nu}$ are in cgs units ($\JUnits$ and $\ergcm3$, respectively).

The optical depth of a cloud is greater at lower frequency.  Therefore, the homogeneous, optically thin case (eq.~\ref{eqn:FFBumpLimitJ} and {eqn:FFBumpLimitu}) is appropriate for the high frequency radio background, while the optically thick case (eq. \ref{eqn:FFBumpLimitJThick} and \ref{eqn:FFBumpLimituThick}) is appropriate for the external lowest frequency radio background.  When there is no scattering, the transition between the two cases occurs when $\tau_{\rm ff} = \alpha_{\rm \nu,ff} R = 1$; for typical values in the IGM, this is
\begin{equation}
\label{eqn:nuCalR}
\nu_{\rm cal} (R) \approx 230\ \Hz\ T_4^{3/2} \delta_e \overline{g_{\rm ff,15}}^{1/2} R_{\rm Mpc}^{1/2}.
\end{equation}
The homogeneous approximation should be valid down to below a kHz for typical IGM clouds, if there is no scattering.  With scattering, the cloud becomes effectively optically thick when $\tau_{\rm eff} \approx \sqrt{\alpha_{\rm \nu,ff} \delta_{\nu}} R = 1$.  If we naively apply the effective mean free path from equation~\ref{eqn:lambdaAbsEff}, we get the following characteristic frequency
\begin{equation}
\label{eqn:nuCalRs}
\nu_{\rm cal,s} (R) \approx 20\ \kHz\ T_4^{-9/34} \delta_e^{12/17} \overline{g_{\rm ff,15}}^{3/17} \Delta_n^{3/17} R_{\rm Mpc}^{6/17} \ell_{0,\rm Mpc}^{-2/17}.
\end{equation}
where $\Delta_n = \mean{\Delta n_e^2} / n_e^2$.  Below this characteristic frequency, the upper bound on $\Delta\nu J_{\nu}$ would go as $\sqrt{\delta_{\nu} / \alpha_{\nu}} \propto \nu^{-5/6}$.  However, as explained in \S~\ref{sec:IGMPropagation}, the approximations used in $\delta_{\nu}$ are unlikely to be appropriate for large scattering optical depths.  

However, I have ignored the time it takes for radio waves to diffuse into the centre of the cloud when scattering is present.  Comparison with the diffusion times derived in \S~\ref{sec:IGMPropagation} shows that the diffusion times are likely to matter at low frequencies if this scattering is present.  In the case with scattering, we must require that the radio waves have enough time to diffuse through the cloud in order to heat it.  Conservatively, I will choose the requirement that the time to diffuse through the cloud radius $R$ is less than or equal to $t_{\rm IGM}$.  From equation~\ref{eqn:tDiff}, this gives us a characteristic frequency:
\begin{equation}
\label{eqn:nuDiff}
\nu_{\rm diff} = 34\ \kHz\ t_{10}^{-3/11} R_{\rm Mpc}^{6/11} \ell_{0,\rm Mpc}^{-2/11} \Delta_n^{3/11} \delta_e^{6/11},
\end{equation}
where $t_{10} = t_{\rm IGM} / (10\ \Gyr)$.  If scattering is present, the free-free absorption bounds do not apply below this frequency.  There may be a way around this problem -- the outer layers of the cloud may disperse as they `evaporate', allowing new radio waves to free stream into the deeper layers of the cloud.    

\begin{table*}
\begin{minipage}{170mm}
\caption{IGM Free-Free Absorption Upper Limits on the sub-MHz Background}
\label{table:FFNuUNuLimits}
\begin{tabular}{lccccccc}
\hline
Phase & $\nu_P$ & $\nu_{\rm cal} (R)^a$ & $\nu_{\rm cal,s} (R)^b$ & $\nu_{\rm diff}^c$ & \multicolumn{3}{c}{Upper Bounds on $\Delta\nu u_{\nu}$} \\
 & & & & & Homogeneous (MHz)$^d$ & Opaque (No scattering)$^e$ & Opaque (Scattering)$^f$ \\
 & ($\Hz$) & ($\Hz$) & ($\kHz$) & ($\kHz$) & ($\erg~\cm^{-3}$) & ($\erg~\cm^{-3}$) & ($\erg~\cm^{-3}$)\\
\hline
Background IGM & $0.20$ & $15$ & $1.2$ & $14$ & $6.7 \times 10^{-12}$ & $2.8 \times 10^{-21}$ & $2.0 \times 10^{-17}~^g$\\
Underdense Ly$\alpha$ forest & $1.4$ & $7.5$ & $2.2$ & $4.8$ & $2.2 \times 10^{-13}$ & $2.4 \times 10^{-23}$ & $2.2 \times 10^{-18}~^g$\\
Ly$\alpha$ forest & $4.5$ & $72$ & $11$ & $17$ & $2.2 \times 10^{-14}$ & $2.4 \times 10^{-22}$ & $5.5 \times 10^{-18}~^g$\\
Low density WHIM & $10$ & $41$ & $18$ & $93$ & $3.2 \times 10^{-10}$ & $1.2 \times 10^{-18}$ & $3.0 \times 10^{-13}~^g$\\
High density WHIM & $45$ & $780$ & $150$ & $480$ & $1.6 \times 10^{-11}$ & $2.4 \times 10^{-17}$ & $1.0 \times 10^{-12}~^g$\\
Cool WHIM & $9.0$ & $870$ & $50$ & $74$ & $3.7 \times 10^{-15}$ & $6.2 \times 10^{-21}$ & $1.9 \times 10^{-17}~^g$\\
Low density cluster & $90$ & $110$ & $120$ & $1900$ & $3.1 \times 10^{-6}$ & $9.5 \times 10^{-14}$ & $1.7 \times 10^{-7}~^g$\\
High density cluster & $280$ & $1100$ & $620$ & $6700$ & $3.1 \times 10^{-7}$ & $9.5 \times 10^{-13}$ & $4.3 \times 10^{-7}~^g$\\
Strong Mg II absorber & $900$ & $2.4 \times 10^5$ & $6200$ & $4800$ & $1.7 \times 10^{-16}$ & $2.7 \times 10^{-17}$ & $1.3 \times 10^{-14}$\\
Dense weak Mg II absorber & $900$ & $26000$ & $2100$ & $2400$ & $5.5 \times 10^{-15}$ & $9.5 \times 10^{-18}$ & $5.1 \times 10^{-14}$\\
Weak Mg II/C IV absorber & $280$ & $26000$ & $1300$ & $1000$ & $5.5 \times 10^{-16}$ & $9.5 \times 10^{-19}$ & $1.8 \times 10^{-15}$\\
C IV/O VI Galaxy Halo & $28$ & $2700$ & $150$ & $200$ & $3.3 \times 10^{-15}$ & $4.9 \times 10^{-20}$ & $1.4 \times 10^{-16}$\\
\hline
\end{tabular}
\\$^a$  Frequency at which the cloud becomes optically thick to free-free absorption, assuming no scattering.\\
$^b$  Frequency at which the cloud becomes effectively optically thick to free-free absorption, assuming the scattering mean free path is given by eq.~\ref{eqn:mfpScattering}.  Note the kHz units.  Assumes $\ell_0 = R$ and $\Delta_n = 1$.\\
$^c$  Frequency at which the diffusion time through the cloud equals $t_{\rm IGM}$, assuming the scattering mean free path is given by eq.~\ref{eqn:mfpScattering}.  Note the kHz units.  Assumes $\ell_0 = R$ and $\Delta_n = 1$.\\
$^d$  Upper bound at MHz if the radio is distributed evenly through the cloud.  The upper bound on $\Delta\nu u_{\nu}$ roughly scales as $\nu^2$, with some logarithmic dependence from $\overline{g_{\rm ff}}$.\\
$^e$  Upper bound on incident external radiation with $\nu < \nu_{\rm cal} (R)$ from cloud evaporation, if the clouds do not scatter radio waves.\\
$^f$  Upper bound on incident external radiation at $\nu = \nu_{\rm cal,s} (R)$ from cloud evaporation, if the cloud's scattering mean free path is given by eq.~\ref{eqn:mfpScattering}, and if the radio has enough time to diffuse through the entire cloud (applies if $\nu_{\rm diff} \le \nu_{\rm cal,s} (R)$).  Below $\nu_{\rm cal,s} (R)$, the energy density bound is weaker, going as $\nu^{-5/6}$.  Eq.~\ref{eqn:mfpScattering} may not be valid at large optical depths.  Assumes $\ell_0 = R$ and $\Delta_n = 1$.\\
$^g$  Since $\nu_{\rm diff} > \nu_{\rm cal,s} (R)$, the radio may not have enough time to diffuse through the entire cloud to evaporate it when $\nu = \nu_{\rm cal,s} (R)$.  Therefore these limits may not apply.
\end{minipage}
\end{table*}

In Table~\ref{table:FFNuUNuLimits}, I have calculated these limits for the various IGM phases considered here.  I have assumed $\Delta_n = 1$ and $\ell_0 = R$ for the case with scattering.  As before, the Lyman-$\alpha$ forest limits are the strongest, because of the low thermal energy density of the Lyman-$\alpha$ forest and its long lifetimes.  Furthermore, the low density of the Lyman-$\alpha$ forest means that it is more transparent, allowing more of the cloud interior to be heated.  Scattering weakens the limits considerably for $\nu \la \nu_{\rm cal,s} (R)$, since the cloud becomes effectively opaque at much higher frequencies than without scattering, shielding the interior from the external radio bath.  Requiring that the diffusion time through the cloud be less than $t_{\rm IGM}$ weakens the limits even more, since then there are no free-free absorption bounds below $\nu_{\rm diff}$.  For $\delta = 1$ Lyman-$\alpha$ forest clouds, this allows large external radio backgrounds below $13\ \kHz$.

The estimates in this section ignore another consideration that may weaken these bounds.  I have assumed that once a layer of the cloud is heated, it disperses and no longer contributes to the opacity.  While their opacity is expected to go down as they heat up, the outer `evaporated' layers of the cloud may still be sufficiently opaque to shield the interior of the cloud from the external radio background.  This will slow down further evaporation when $\tau \gg 1$, which will especially be a problem when scattering is strong.

In this section I have only considered bounds from the heating of the entire cloud.  There may other, more subtle statements that can be made.  For example, even if the sub-MHz radio cannot penetrate deep into the interior of the cloud, it can still heat the exterior of the cloud, and cause it to become more highly ionized.  The heated layers would then be visible in absorption lines like C IV or O VI at the same velocity as the cloud interior.  The number statistics of highly ionized absorbers compared to lower ionization absorbers at the same velocity may then set limits on the sub-MHz background even in the case of strong scattering and high optical depth.

\subsection{Luminosity Density Limits}
\label{sec:FFEpsilonBound}
We can also set limits on the \emph{luminosity} density of low frequency radio sources within the IGM.  The energy density of a region is simply the luminosity in that region integrated over the time the photons could have lasted in the IGM: $u_{\nu} = \epsilon_{\nu} t_{\nu}$.  

\emph{Homogeneous case} -- In the case when the sources are distributed inside the cloud continuously (equation~\ref{eqn:JFFLimitInt}), the limit on the luminosity density is
\begin{equation}
\Delta\nu \epsilon_{\nu} \la \frac{2 nkT}{c t_{\rm IGM} \alpha_{\nu,\rm ff} t_{\nu}}.
\end{equation}
Suppose that free-free absorption is the main source of absorption, with synchrotron absorption negligible.  As discussed in \S~\ref{sec:IGMPropagation}, this is probably a good assumption for most of the IGM; even in clusters, it will hold for radiation with frequencies below the Razin frequency.  If we ignore other sources of absorption, $\alpha_{\nu} = \alpha_{\nu, \rm ff}$.  We can define an absorption time-scale for that particular parcel of gas as $t_{\rm abs} = (c \alpha_{\nu})^{-1}$, which is the time-scale that a photon of frequency $\nu$ that is trapped \emph{within that gas parcel} will be absorbed.  This gives us
\begin{equation}
\Delta\nu \epsilon_{\nu} \la \frac{2 nkT}{t_{\rm IGM}} \frac{t_{\rm abs}}{t_{\nu}}.
\end{equation}

The choice of $t_{\nu}$ is potentially very complicated, since it is spatially averaged.  If the photons are trapped in a homogeneous IGM phase, $t_{\nu} = \min [\mean{t_{\rm abs}}, t_{\rm IGM}]$, where $\mean{t_{\rm abs}} = \mean{c \alpha_{\nu}}^{-1}$ is the mean absorption time-scale for a photon of frequency $\nu$ over the path it travelled.  

In the spatially homogeneous opaque limit with only free-free absorption, when $t_{\nu} = \mean{t_{\rm abs}} = t_{\rm abs} = t_{\rm ff}$, we enter a simple `calorimeter' limit:
\begin{equation}
\Delta\nu \epsilon_{\nu} \la \frac{2 nkT_{\rm IGM}}{t_{\rm IGM}},
\end{equation}
which can be evaluated as\footnote{Note that $1\ \erg\ \cm^{-3}\ \sec^{-1} = 7.7 \times 10^{39} \Lsun \Mpc^{-3}$.}
\begin{equation}
\Delta\nu \epsilon_{\nu} \la 3.4 \times 10^5 \Lsun \Mpc^{-3} \delta_e T_4 \left(\frac{t_{\rm IGM}}{\rm Gyr}\right)^{-1}.
\end{equation}
This equation simply says that all of the radio emission in a region heats up the local IGM, and the luminosity integrated over time cannot exceed the thermal content of the IGM.  The ideal properties of an IGM parcel as a radio detector are obvious: it must be relatively low density and cold, and it must last a long time.  However, the frequency at which the IGM becomes a radio calorimeter increases with density.  Low density IGM phases are good detectors at low frequency, while high density IGM phases are better for high frequency.  In the case when other sources of absorption are present, the true limit of $\Delta\nu \epsilon_{\nu}$ is larger by a factor of $t_{\rm ff}/t_{\rm abs}$, since there are other sinks for the luminosity.

The exact frequency $\nu_{\rm cal}$ at which an IGM cloud becomes calorimetric depends on the radiative transfer within the cloud.  The longest possible time a radio photon can have to be absorbed (thus heating the cloud) is $t_{\rm IGM}$, since the cloud has a finite age.  This case applies when scattering traps radiation effectively.  The absorption time equals $t_{\rm IGM}$ at the frequency
\begin{equation}
\label{eqn:nuCalctIGM}
\nu_{\rm cal} (ct_{\rm IGM}) = 13\ \kHz\ \delta_e \overline{g_{\rm ff,15}}^{1/2} t_{10}^{1/2} T_4^{-3/4}
\end{equation}
The shortest time a radio photon can have to cross the cloud and escape is $R / c$, which applies when radio emission free-streams out of the cloud.  The frequency when the absorption time is $R / c$ is simply the frequency at which it would become opaque with no scattering\footnote{Since we care about the \emph{absorption time} and not the \emph{displacement from the source before absorption}, the optical depth we care about is the true absorption optical depth, not the effective absorption optical depth.}, $\nu_{\rm cal} (R)$, given in equation~\ref{eqn:nuCalR}.  I calculate the frequencies when $t_{\rm ff}$ equals each of these times in Table~\ref{table:IGMasFFCalorimeter}.  

\begin{table*}
\begin{minipage}{140mm}
\caption{The IGM as a Free-Free Calorimeter}
\label{table:IGMasFFCalorimeter}
\begin{tabular}{lccccc}
\hline
Phase & $R$ & $\nu_P$ & $\nu_{\rm cal} (R)^a$ & $\nu_{\rm cal} (ct_{\rm IGM})^b$ & $\nu \epsilon_{\nu}$ Upper Bound$^c$\\
 & ($\Mpc$) & ($\Hz$) & ($\Hz$) & ($\kHz$) & ($\Lsun \Mpc^{-3}$)\\
\hline
Background IGM & $1000$ & $0.2$ & $15$ & $0.026$ & $67$\\
Underdense Ly$\alpha$ forest & $1$ & $1.4$ & $7.5$ & $0.93$ & $5600$\\
Ly$\alpha$ forest cloud & $1$ & $4.5$ & $72$ & $8.9$ & $56000$\\
Low density WHIM & $1$ & $10$ & $41$ & $1.7$ & $2.8 \times 10^7$\\
High density WHIM & $1$ & $45$ & $780$ & $31$ & $5.6 \times 10^8$\\
Cool WHIM & $1$ & $9.0$ & $870$ & $42$ & $1.5 \times 10^5$\\
Low density cluster$^d$ & $1$ & $90$ & $110$ & $1.5$ & $2.3 \times 10^{12}$\\
High density cluster$^d$ & $1$ & $280$ & $1100$ & $14$ & $2.3 \times 10^{13}$\\
Strong Mg II absorber & $0.001$ & $900$ & $2.4 \times 10^5$ & $17000$ & $6.7 \times 10^{11}$ \\
Dense weak Mg II absorber & $10^{-5}$ & $900$ & $26000$ & $3100$ & $2.3 \times 10^{13}$\\
Weak Mg II/C IV absorber & $0.001$ & $280$ & $26000$ & $3100$ & $2.3 \times 10^{10}$\\
C IV/O VI Galaxy Halo & $0.1$ & $28$ & $2700$ & $140$ & $1.4 \times 10^7$\\
\hline
\end{tabular}
\\$^a$  Frequency at which the mean free absorption path is the width of the IGM phase, the minimum possible value for $\nu_{\rm cal}$ (if only free-free absorption is present).\\
$^b$  Frequency at which the free-free absorption time-scale is $t_{\rm IGM}$, the maximum possible value for $\nu_{\rm cal}$.\\
$^c$  Upper bound on the luminosity density within each cloud for $\nu \le \nu_{\rm cal}$.  Assumes that the luminosity is distributed continuously throughout the cloud, so that no part of the cloud is shielded, and that there are no other sources of absorption.\\
$^d$  Synchrotron absorption may exceed free-free absorption, making the luminosity density bound invalid for $\nu > \nu_{R}$.
\end{minipage}
\end{table*}

The calorimetric limit applies for all frequencies $\nu_P \le \nu \le \nu_{\rm cal}$, where $\nu_{\rm cal}$ is between $\nu_{\rm cal} (t_{\rm IGM})$ and $\nu_{\rm cal} (R/c)$.  In Table~\ref{table:IGMasFFCalorimeter}, I calculate the calorimetric limits on the luminosity density of sub-MHz radio for each phase, assuming free-free absorption is the only source of absorption.  The Lyman-$\alpha$ forest sets very good limits on the luminosity density of the sub-MHz Universe.  The existence of $\delta = 1$ Lyman-$\alpha$ clouds implies that $\Delta\nu \epsilon_{\nu} \la 6\times 10^4\ \Lsun\ \Mpc^{-3}$ for frequencies $\nu_P \la \nu \la \nu_{\rm cal}$.  For comparison, this is smaller than the gamma-ray emissivity of the Universe \citep{Coppi97}, and only a tiny fraction of the bolometric emissivity of the Universe.  Denser and hotter clouds, like WHIM clouds or metal line absorbers set relatively poor limits on the sub-MHz luminosity density.

\emph{Optically thin case} -- If the Universe is optically thin to a radio photon, so that the mean absorption time-scale along its path length is much longer than the age of the Universe, however, then $t_{\nu} = t_H$ since the Universe has been filling up with radio emission for its entire history.  Note that the condition $\mean{t_{\rm abs}} \gg t_H$ must account for the fact that the radio photon may traverse different IGM phases with different \emph{local} absorption time-scales $t_{\rm abs}$.  Again, assuming only free-free absorption is present, we have the condition
\begin{eqnarray}
\Delta\nu \epsilon_{\nu} & \la & \frac{2nkT}{t_{\rm IGM}} \frac{t_{\rm abs}}{t_H}\\
 & \la & 1.6 \times 10^{10}\ \Lsun \Mpc^{-3} \nu_{\rm MHz}^2 \delta_e T_4^{5/2} \left[\frac{t_{\rm IGM}}{\rm Gyr} \frac{t_H}{10\ \rm Gyr}\right]^{-1},
\end{eqnarray}
where again, $T_4 = T / (10^4\ \Kelv)$.  These bounds are naturally much weaker, since the IGM is transparent to radio waves at high enough frequency, and the radio waves keep most of their energy.

\section{Radiation Pressure on IGM Clouds}
\label{sec:CloudCrush}
No radio waves below the plasma frequency can traverse an IGM cloud.  Since most of the IGM by volume is extremely low density ($\delta < 0.1$) \citep{Bi97}, this means that most of the IGM could be filled with extremely low frequency waves ($\nu \la \Hz$) which cannot enter the IGM clouds that serve as our detectors in \S~\ref{sec:FFAbsorption}.  These radio waves would simply bounce off the clouds without being absorbed.  However, the bounce itself is an impulse on the cloud that might have detectable effects.  Thus, the radiation pressure of radio waves below the plasma frequency can \emph{squeeze} IGM clouds, precisely \emph{because} they reflect off them.

In fact, the radiation pressure of radio waves can also squeeze the IGM clouds if radio waves are effectively absorbed by \citep{Ciotti04} or scattered within them.  Even without scattering, radio waves will be absorbed by free-free absorption for frequencies lower than $\nu_{\rm cal} (R)$ (eq.~\ref{eqn:nuCalR}).  Scattering can only increase the coupling between radio waves and the cloud.  If the naive scattering mean free paths in equation~\ref{eqn:mfpScattering} are used, an IGM cloud is optically thick to scattering below a frequency
\begin{equation}
\label{eqn:nuS}
\nu_S \approx 220\ \kHz\ R_{\rm Mpc}^{3/11} \Delta_n^{3/11} \delta_e^{6/11} \ell_{0,\rm Mpc}^{-2/11}.
\end{equation}
I conservatively assume that there is no scattering, so that the maximum frequency $\nu_{\rm int}$ at which a cloud is optically thick to either absorption or scattering simply equals $\nu_{\rm cal} (R)$ (eq.~\ref{eqn:nuCalR}).

There are two characteristic speeds for the collapse.  The first is the sound speed of the IGM (c.f. eq.~\ref{eqn:tSound}), which is
\begin{equation}
c_s \approx \sqrt{\frac{10 k T}{3 m_H}} \approx 1.7 \times 10^6\ \cm\ \sec^{-1} T_4^{1/2}.
\end{equation}
A relatively small pressure imbalance will cause the cloud to collapse, essentially quasi-statically and adiabatically, until its internal pressure is enough to resist the imbalance.  If the sound crossing time is small compared to $t_{\rm IGM}$, then the pressure imbalance would have forced an IGM cloud to have a structure different than the one we observe.  If $R / c_s < t_{\rm IGM}$, the external sub-$\nu_P$ radiation pressure $P_{\rm rad}$ cannot be much greater than the thermal pressure of the cloud $P_c \approx 2 n_e k T$, or 
\begin{equation}
\label{eqn:SoundURad}
u_{\rm rad} \la 2.0 \times 10^{-18} \ergcm3\ \delta_e\ T_4.
\end{equation}  

When the external pressure is much greater than the cloud pressure, however, the collapse can occur much more quickly.  Consider what would happen at the edge of a cloud immersed in a sub-$\nu_{\rm int}$ radiation bath with an extremely large pressure $P_{\rm rad}$.  The radiation pressure will rapidly accelerate the particles in the edge of the cloud until they exceed the local sound speed.  A shock will form in the edge of the cloud.  The shocking will continue until the local pressure balances the external pressure.  Either the post-shocked gas will continue interacting with the sub-$\nu_{\rm int}$ radiation, in which case the exterior radiation pressure will continue to drive it inwards, or it will be transparent, in which case the radiation will now shock the next layer in until the entire cloud is shocked.

Therefore, a shock will propagate inwards through the cloud, heating it until the internal pressure equals the external pressure $P_{\rm rad}$.  The Rankine-Hugoniot conditions for a shock imply
\begin{equation}
\frac{P_2}{P_1} = 1 + \frac{2 \gamma_g}{\gamma_g + 1} (M_1^2 - 1),
\end{equation}
where $P_1$ is the pre-shock pressure, $P_2$ is the post-shock pressure -- equal to $P_{\rm rad}$ in this case, $M_1 = v / c_{s,1}$ is the shock Mach number through the unshocked gas, and $\gamma_g$ is $5/3$ for an ionized monatomic gas.  We find that
\begin{equation}
P_{\rm rad} = \frac{3}{4} \rho_1 v^2 - \frac{1}{4}P_1
\end{equation}
In the case when $P_{\rm rad} = P_1$, we simply get $v = c_{s,1}$ as expected.  Since the IGM clouds we do see have not been destroyed yet, we can suppose that $v \la R / t_{\rm IGM}$.  Using the fact that $u_{\rm rad} = 3 P_{\rm rad}$, we find that
\begin{equation}
\label{eqn:ShockedURad}
u_{\rm rad} \la \left[9.0 \times 10^{-17} \delta_e R_{\rm Mpc}^2 t_{10}^{-2} - 2.6 \times 10^{-19} \delta_e T_4\right] \ergcm3
\end{equation}  
where $t_{10} = t_{\rm IGM} / {\rm 10\ Gyr}$, $R_{\rm Mpc} = R / {\rm Mpc}$, and $T_4 = T / (10^4 K)$ for the initial cloud.  

If the post-shock region is radiating efficiently, the shock may instead be isothermal, with the final temperature equal to the initial temperature.  In this case
\begin{equation}
\frac{P_3}{P_1} = M_1^2,
\end{equation}
where $P_3$ is the pressure of the radiatively cooled region.  If we now assume that $P_3 = P_{\rm rad}$, we find that
\begin{equation}
\label{eqn:IsothermShockURad}
u_{\rm rad} \la 7.2 \times 10^{-17} \delta_e R_{\rm Mpc}^2 t_{10}^{-2} \ergcm3.
\end{equation}
In all considered cases, the limits on $u_{\rm rad}$ from the isothermal case are not as conservative as those from the non-radiating case.

\begin{table*}
\begin{minipage}{170mm}
\label{table:CloudCrush}
\caption{Radiation Pressure Constraints on sub-$\nu_{\rm int}$ backgrounds.}
\begin{tabular}{lcccccc}
\hline
Phase & $\nu_{\rm cal} (R)$ & $\nu_s (R)^a$ & $P_{\rm therm}^b$ & $R / t_{\rm IGM}$ & $c_s$ & $u_{\rm rad} (\nu \le \nu_{\rm int})$ Upper Bound$^c$\\
 & ($\Hz$) & ($\kHz$) & ($\erg\ \cm^{-3}$) & ($\kms$) & ($\kms$) & ($\erg\ \cm^{-3}$)\\
\hline
Background IGM & $15$ & $14$ & $1.4 \times 10^{-21}$ & $98000$ & $17$ & $1.8 \times 10^{-13}$\\
Underdense Ly$\alpha$ forest & $7.5$ & $52$ & $6.9 \times 10^{-20}$ & $160$ & $17$ & $2.1 \times 10^{-19}$\\
Ly$\alpha$ forest cloud & $72$ & $180$ & $6.9 \times 10^{-19}$ & $160$ & $17$ & $2.1 \times 10^{-18}$\\
Low density WHIM & $41$ & $540$ & $3.5 \times 10^{-16}$ & $160$ & $170$ & $1.0 \times 10^{-15}$\\
High density WHIM & $780$ & $2800$ & $6.9 \times 10^{-15}$ & $160$ & $170$ & $2.1 \times 10^{-14}$\\
Cool WHIM & $870$ & $480$ & $2.8 \times 10^{-18}$ & $11$ & $17$ & $4.4 \times 10^{-16}$\\
Low density cluster & $110$ & $5900$ & $2.8 \times 10^{-12}$ & $1600$ & $1700$ & $8.3 \times 10^{-12}$\\
High density cluster & $1100$ & $2.1 \times 10^4$ & $2.8 \times 10^{-11}$ & $1600$ & $1700$ & $8.3 \times 10^{-11}$\\
Strong Mg II absorber & $2.4 \times 10^5$ & $3.9 \times 10^4$ & $2.8 \times 10^{-14}$ & $49$ & $17$ & $8.8 \times 10^{-13}$\\
Dense weak Mg II absorber & $26000$ & $2.5 \times 10^4$ & $2.8 \times 10^{-14}$ & $16$ & $17$ & $8.3 \times 10^{-14}$\\
Weak Mg II/C IV absorber & $26000$ & $1.1 \times 10^4$ & $2.8 \times 10^{-15}$ & $16$ & $17$ & $8.3 \times 10^{-15}$\\
C IV/O VI Galaxy Halo & $2700$ & $1400$ & $2.8 \times 10^{-17}$ & $98$ & $17$ & $3.6 \times 10^{-15}$\\
\hline
\end{tabular}
\\$^a$  Frequency at which the cloud becomes optically thick to scattering, if eq.~\ref{eqn:mfpScattering} describes the scattering mean free path and $\Delta_n = 1$ and $\ell_0 = R$.\\
$^b$  Pressure of the cloud; if $R / t_{\rm IGM} \le c_s$, this is the limit on the external sub-$\nu_P$ radiation pressure.\\
$^c$  Note that $u_{\rm rad} = 3 P_{\rm rad}$.
\end{minipage}
\end{table*}

Equation~\ref{eqn:ShockedURad} holds only if $M_1 \ge 1$, because of the Second Law of Thermodynamics; in the case when $R / t_{\rm IGM} < c_{s,1}$, equation~\ref{eqn:SoundURad} is appropriate since pressure imbalances will be propagated at least at the sound speed.   However, since $t_{\rm IGM}$ was chosen to be at most $t_{\rm sound}$ in \S~\ref{sec:IGMPhases}, eq.~\ref{eqn:ShockedURad} is always the appropriate choice.

Note that the radiation pressure bound is \emph{bolometric} for all electromagnetic radiation below $\nu_{\rm int}$.  \emph{All} incident radio waves with frequencies below this frequency will be reflected off, scattered by, or absorbed by the cloud, and will therefore squeeze the cloud.  Only if the wavelength of the incident waves are greater than the size of the cloud will the radiation fail to interact, and the background plasma density with a low enough plasma frequency for a radio wave with wavelengths of kpc to Mpc to exist is inconceivably low.  

\section{Synchrotron Heating in Clusters}
\label{sec:SynchHeat}
Cluster gas is very hot and dense, so the limits I derived from free-free absorption (\S~\ref{sec:FFUBound}) and radiation pressure (\S~\ref{sec:CloudCrush}) for the sub-MHz radio background in galaxy clusters are weak.  However, there is another gas phase in galaxy clusters that can serve as a sub-MHz detector: the relativistic phase.  Some clusters are filled with high energy cosmic ray electrons and possibly positrons (CRs), which emit synchrotron radio emission observed at MHz to GHz frequencies \citep{Ferrari08}.  These electrons also Inverse Compton scatter CMB photons to produce hard X-ray emission \citep{Rephaeli08}.  Synchrotron cooling and IC cooling off the CMB should dominate the cooling processes at high energies.  Inverse Compton losses off starlight produced by the cluster are relatively weak; given a typical cluster luminosity of $L \approx 10^{46}\ \ergps$ and a radius of $R \approx 1\ \Mpc$, the typical starlight energy density is $U_{\star} = L / (4 \pi R^2 c) \approx 3 \times 10^{-15}\ \erg\ \cm^{-3}$.  Indeed, cluster radio emission is observed to fall off steeply, as would be expected by synchrotron and IC cooling. 

However, a large radio background can actually \emph{accelerate} particles, especially at lower energies, through \emph{synchrotron absorption}.  This effect can overcome Inverse Compton and synchrotron cooling in certain circumstances \citep*{Ghisellini88}.  The synchrotron heating rate is
\begin{equation}
\label{sec:bHeat}
-b_{\rm heat} \approx \frac{c^2}{2} E^2 \frac{\partial}{\partial E} \left[\frac{N(E)}{E^2}\right] N(E)^{-1} \int_0^{\infty} \frac{I_{\nu}}{\nu^2} P(\nu, E) d\nu
\end{equation}
Synchrotron heating will create a rising CR electron/positron spectrum at low energies that peaks at some frequency depending on the IC and synchrotron cooling time-scales.  Therefore, the CR electrons and positrons in a galaxy cluster also serve as a radio detector.  The very fact that no such peak is seen in the radio spectra of clusters places limits on the sub-MHz radio background in these regions.  

\begin{figure}
\centerline{\includegraphics[width=8cm]{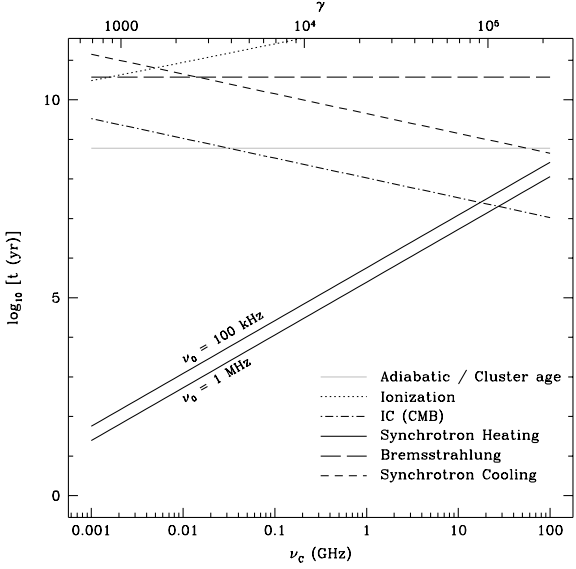}}
\caption{Cooling and heating times for CR $e^{\pm}$ in a galaxy cluster with a magnetic field strength of $0.5~\muGauss$ and a gas density of $10^{-3}~\cm^{-3}$.  For the synchrotron heating times, I assume each $\Delta\nu_0 u_{\nu} (\nu_0)$ is the free-free absorption upper limit on a homogeneous background filling the cluster gas (\S~\ref{sec:FFUBound}), for $\nu_0 = 100\ \kHz$ and $1\ \MHz$.  Although synchrotron heating by $100\ \kHz$ photons is more effective than by MHz photons at fixed $\Delta\nu_0 u_{\nu} (\nu_0)$, the upper limit on the 100 kHz background is much more severe, so the synchrotron heating allowed by the free-free absorption bound is weaker.\label{fig:ClusterLosses}}
\end{figure}

There is a competition between heating and cooling (including Inverse Compton cooling off the low frequency background itself) at every CR energy.  In order for synchrotron heating to be effective, the absorbed radiation frequency $\nu_0$ must be greater than the Razin frequency $\nu_R$ (eq.~\ref{eqn:nuRazin}), and the synchrotron heating time-scale must be less than the cooling time-scale from all other processes: $t_{\rm heat} < [t_{\rm synch}^{-1} + t_{\rm IC, CMB}^{-1} + t_{\rm IC, sub-MHz}^{-1} + t_{\rm IGM}^{-1}]^{-1}$.  The synchrotron cooling time-scale of electrons/positrons observed in synchrotron radio at $\nu_C = \nu_{\rm C, GHz} \GHz$ is
\begin{equation}
t_{\rm synch} = 1.59 \times 10^9~\yr~\nu_{\rm C, GHz}^{-1/2} B_{\rm \mu G}^{-3/2}.
\end{equation}
The Inverse Compton cooling time-scale from low frequency photons at frequency near $\nu_0$ is
\begin{equation}
t_{\rm IC} = 6.32 \times 10^{-5}~\yr~\nu_{\rm C, GHz}^{-1/2} B_{\rm \mu G}^{1/2} [\Delta\nu_0 u_{\nu}(\nu_0)]^{-1},
\end{equation}
where $\Delta\nu_0 u_{\nu} (\nu_0)$ is in cgs ($\ergcm3$).  Additional Inverse Compton cooling comes from the CMB:
\begin{equation}
t_{\rm IC, CMB} = 1.50 \times 10^8~\yr~\nu_{\rm C, GHz}^{-1/2} B_{\rm \mu G}^{1/2}.
\end{equation}
The synchrotron heating time-scale off photons near frequency $\nu_R \la \nu_0 \ll \nu$, if the electron spectrum goes as\footnote{There is some residual dependence on the electron spectrum shape, because the effectiveness of synchrotron absorption is affected by stimulated synchrotron emission.  Using an $E^{-2}$ or $E^{-4}$ CR electron/positron spectrum will not alter these estimates much.} $E^{-3}$, is
\begin{equation}
t_{\rm heat} = 1.93 \times 10^{-7}~\yr~\nu_{\rm C, GHz}^{4/3} \nu_{\rm 0,kHz}^{5/3} B_{\rm \mu G}^{-2} [\Delta\nu_0 u_{\nu}(\nu_0)]^{-1}.
\end{equation}
As seen in Figure~\ref{fig:ClusterLosses}, if the sub-MHz radio background is the maximum allowed by the free-free absorption bounds, synchrotron heating is by far the most important process at low energies for a galaxy cluster.  The fact that we do not see a turnover in the radio synchrotron spectrum of galaxy clusters at $\sim 10\ \GHz$ implies that the sub-MHz radio background in clusters must be far smaller than that allowed by the free-free absorption bounds.  Using these time-scales gives us the following condition for synchrotron heating to be too weak to alter the spectrum:
\begin{equation}
\label{eqn:SynchHeatingBound}
\Delta\nu_0 u_{\nu} (\nu_0) < 10^{-16} \ergcm3\ \frac{1.2 B_{\rm \mu G}^{2} + 13 + 1.9 B_{\rm \mu G}^{1/2} t_9^{-1} \nu_{\rm C, GHz}^{1/2}}{\nu_{\rm C, GHz}^{-11/6} \nu_{\rm 0,kHz}^{-5/3} B_{\rm \mu G}^{5/2} - 0.000305}
\end{equation}
where $t_9 = t_{\rm IGM} / \Gyr$.  

\begin{figure}
\centerline{\includegraphics[width=8cm]{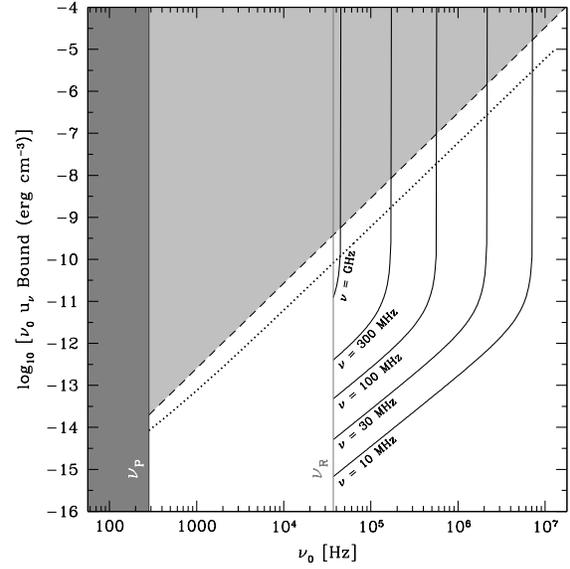}}
\caption{Limits on the homogeneous sub-MHz energy densities in a cluster with $n = 10^{-3}\ \cm^{-3}$, $B = 0.5\ \muGauss$, $T = 10^8~\Kelv$, and $t_{\rm IGM} = 0.6~\Gyr$.  The dashed line (\emph{light grey shading}) is the (optically thin) free-free absorption limit (eq.~\ref{eqn:FFBumpLimitu}).  Solid lines are synchrotron heating limits, assuming that there is no spectral downturn at synchrotron frequencies of 10 MHz to GHz.  Synchrotron reheating limits are generally much stronger than the free-free absorption limits, but only apply above the Razin cutoff ($\nu_R$) and require a low frequency radio detection of the cluster.  The Inverse Compton limits (dotted) for a cluster with a Coma cluster-like CR electron spectrum are the strongest constraints below the Razin cutoff.\label{fig:ClusterLimits}}
\end{figure}

In practice, the synchrotron heating bound appears three-sided, as seen in Figure~\ref{fig:ClusterLimits}.  Synchrotron heating by low frequency radio waves with $\nu_0 < \nu_R$ is not effective because of the Razin cutoff.  At high CR electron/positron energies, Inverse Compton cooling off the low frequency radio background becomes more effective than synchrotron heating from those same photons.  Then synchrotron heating ceases to be important.  The target photon frequency $\nu_0$ where this happens for electrons/positrons observed at $\nu_{\rm C, GHz} \GHz$ is:
\begin{equation}
\nu_0 = 32 \kHz\ \nu_{\rm C, GHz}^{-11/10} B_{\rm \mu G}^{3/2}.
\end{equation}
This is the high-frequency cutoff for the excluded region visible in Figure~\ref{fig:ClusterLimits}.  Finally, for intermediate frequencies $\nu_0$, the synchrotron heating must primarily compete with IC cooling off the CMB, and to a lesser extent, synchrotron cooling and adiabatic losses.  We see from equation~\ref{eqn:SynchHeatingBound} that the bounds on $\Delta\nu_0 u_{\nu} (\nu_0)$ in this regime go as $\nu_0^{5/3}$, as the shown for bottom of the excluded region in Figure~\ref{fig:ClusterLimits}.

For a galaxy cluster with microGauss magnetic fields, the upper limit on radio emission above the Razin cutoff is very small from the lack of a spectral peak at GHz:
\begin{equation}
\nu_0 u_{\nu} (\nu_0) \la 10^{-15} \ergcm3\ \nu_{\rm C, GHz}^{11/6} \nu_{\rm 0,kHz}^{5/3} B_{\rm \mu G}^{-5/2}.
\end{equation}
The synchrotron spectrum of the Coma cluster has already been observed down to $30\ \MHz$, and the limits on its sub-MHz energy density are strong.  LOFAR, which can observe all the way down to 15 MHz, can place incredibly strong bounds on kHz to MHz radio emission.  For example, synchrotron heating from 100 kHz radio would alter the radio spectrum at these frequencies as long as $\nu_0 u_{\nu} (\nu_0) \ga 1 \times 10^{-15}\ \ergcm3\ B_{\rm \mu G}^{-5/2}$.  This is roughly comparable to the starlight energy density of a galaxy cluster; recall that, unlike starlight, sub-MHz radio emission might be trapped for a long time in clusters by plasma scattering, so that a very low luminosity can accumulate for a long time to give a large energy density.

The synchrotron absorption argument should also apply wherever there are strong enough magnetic fields and observable cosmic ray electrons.  For example, it should also apply to structure formation shocks, which are believed to accelerate cosmic rays and produce radio emission.  

In addition to synchrotron absorption, there should be transition absorption \citep{Fleishman89}.  In analogy with synchrotron heating, we would expect a transition heating effect, where low energy particles are heated by transition absorption.  I calculated the transition heating time-scale for an $E^{-3}$ CR electron/positron spectrum, using equation \ref{sec:bHeat}.  Unlike the synchrotron reheating times, transition absorption is inefficient in clusters:
\begin{equation}
t_{\rm heat} = 4.5 \times 10^6~\yr~B_{\rm \mu G}^{-1} \nu_{\rm C, GHz} \nu_{\rm 0,kHz}^{7/3} n_{-3}^{1/3} \ell_{\rm 0,Mpc}^{2/3} \Delta_n^{-1} [\nu_0 u_{\nu}(\nu_0)]^{-1}.
\end{equation}
Therefore transition absorption provides no strong constraints on the sub-MHz background in clusters.

\section{Inverse Compton Bounds in Clusters}
\label{sec:ICBound}
The CR population of electrons in galaxy clusters should also Inverse Compton scatter any sub-MHz radio emission in the cluster to much higher frequencies, such as MHz to GHz radio.  Therefore the MHz to GHz radio emission of the cluster limits the amount of sub-MHz radio of the cluster.

The Inverse Compton emissivity of photons of energy $E_{\gamma}$ from an electron/positron spectrum $N_e(E_e)$ scattering a photon field with number density $n_{\rm ph} (\epsilon)$ at original photon energy $\epsilon$ is given by \citep[e.g.,][]{Schlickeiser02} as
\begin{equation}
Q_{\gamma} (E_{\gamma})_{\rm IC} = \int_0^{\infty} n_{\rm ph} (\epsilon) d\epsilon \int_{E_{\rm min}}^{\infty} \frac{d\sigma (E_{\gamma}, \epsilon, E_e)}{dE_{\gamma}} c N_e (E_e) dE_e,
\end{equation}
in photons per unit volume per unit time per unit energy.  The cross section in this formula is
\begin{equation}
\frac{d\sigma}{dE_\gamma} = \frac{3 \sigma_T}{4 \epsilon \gamma^2} (2q\ln q + 1 + q - 2 q^2)
\end{equation}
and $q \approx E_{\gamma} / (4 \gamma^2 \epsilon)$ in the Thomson limit.  The Inverse Compton emissivity at frequency $\nu_1 = E_{\gamma} / h$ is converted into a specific luminosity as $L_{\nu} = h^2 \nu_1 V Q_{\gamma} (E_{\gamma})$.  The condition that $L_{\nu} \la L_{\nu, \rm obs}$ at frequency $\nu_1$ then becomes
\begin{equation}
u_{\nu} (\nu_0) \la \frac{4 L_{\nu, \rm obs}}{3 \sigma_T c V} \frac{\nu_0}{\nu_1} \left[\int_{\gamma_{\rm min}}^{\infty} (2q\ln q + 1 + q - 2 q^2) \frac{dN}{d\gamma} \gamma^{-2} d\gamma \right]^{-1}.
\label{eqn:ICBound}
\end{equation}

In Figure~\ref{fig:ClusterLimits}, I show the Inverse Compton bounds on the sub-MHz emission in the Coma Cluster (\emph{dotted}), using the CR electron spectrum derived from the synchrotron MHz to GHz radio spectrum (\S~\ref{sec:IGMPropagation}).  Near the plasma frequency, the IC bound is $8 \times 10^{-15}\ \erg\ \cm^{-3}$, which is somewhat greater than the energy density in starlight in a cluster.  At higher frequencies, the IC bound grows as $\gamma^{-2}$.  An electron that emits 30 MHz synchrotron radiation ($\gamma \approx 1000$) can boost kHz photons to 4 GHz photons, the highest frequency radio emission seen from the Coma cluster.  When $\nu_0 \ga \kHz$, the integral over $dN/d\gamma$ is taken to be constant, since we do not know the spectrum at lower energies, and every CR electron/positron observed in synchrotron is energetic enough to boost the $\nu_0$ frequency photons to observable photons of frequency $\nu_1$.\footnote{The $(2q\ln q + 1 + q - 2 q^2)$ in the integral of eq.~\ref{eqn:ICBound} is of order $\sim 1$ for $\nu_0 \ga \kHz$ for all CR electrons/positrons observed at $30\ \MHz \le \nu_1 \le 4\ \GHz$.}  Thus the bound on $u_{\nu} (\nu_0)$ goes as $\nu_0^2$.  If the low energy electron spectrum is ever known better, that would increase the strength of the bounds at high $\nu_0$.  

We see that the IC bounds are stronger than the free-free absorption bound (\emph{dashed}), which were weak because cluster plasma is dense and hot.  The Inverse Compton constraints are not as strong as the synchrotron heating constraints (\emph{solid}) in a cluster like Coma, where data below 100 MHz is available.  Unlike the synchrotron heating constraints, the Inverse Compton bound extends below the Razin cutoff.

LOFAR will be able to measure the radio spectrum of several clusters down to $\sim 15\ \MHz$.  As with the synchrotron heating bounds, these radio observations will allow us to set strong limits on the sub-MHz radio emission in several clusters by the Inverse Compton argument.

\section{Potential Limits from Ultra High Energy $\gamma$-rays}
\label{sec:UHELimits}
If two photons each have an energy greater than $m_e c^2$ in their centre of mass frame, the pair of photons can convert into an electron-positron pair.  For photons of energy $E_1$ and $E_2$ in the observer frame, the pair-production condition is that $E_1 E_2 \ga m_e^2 c^4$.  Pair production is expected to be an important source of opacity for PeV photons as they interact with the CMB \citep{Moskalenko06}.  A large radio background can also serve as a source of opacity for ultra high energy (UHE) photons.  For a MHz radio photon, however, the target photons must have energy $2 \times 10^{20}$ eV, and for kHz radio, the target photons must be at least $2 \times 10^{23}$ eV.

No UHE photons have ever been detected, but if extragalactic photons of high enough energy were ever detected, they would provide potent constraints on the sub-MHz radio background.  The absorption length scale for pair production is $\lambda_{\gamma\gamma} \approx (n_{\gamma} \sigma_T)^{-1}$.  The path length is 1 Mpc only when $n_{\gamma} \approx 0.5\ \cm^{-3}$ or $u \approx 3.2 \times 10^{-24} \nu_{\rm kHz}\ \erg\ \cm^{-3}$.  For comparison, the $\delta = 1$ Ly$\alpha$ forest free-free absorption bound implies the sub-MHz radio photon density is $n_{\gamma} \approx 3300\ \nu_{\rm kHz} \cm^{-3}$ even in the optically thin, homogeneous case.  Thus detections of extragalactic photons with energies above $10^{20}$ eV would increase the bounds on the extragalactic sub-MHz radio background by several orders of magnitude.  

Pair production limits would have the advantage of being relatively more powerful at higher frequency, as opposed to the free-free absorption bound which is more powerful at lower frequency.  It is more likely that UHE photons with lower energy will be detected if their spectrum is falling, which in turn requires higher energy radio photons for pair production to happen.  Indeed, there probably is some flux of photons of energy $10^{20 - 21}\ \eV$ produced by the Greiden-Zatsepin-Kuzmin process \citep[e.g.,][]{Wdowczyk72,Gelmini08}, while there is no obvious mechanism for making photons of much higher energy than that.  Also, the pair production path length scales inversely with the number density of photons, or $\lambda_{\gamma \gamma} \propto (\Delta\nu u_{\nu})^{-1} \nu$, whereas the free-free absorption path length scales as $\lambda_{\rm ff} \propto (\Delta\nu u_{\nu})^{-1} \nu^2$.

Unfortunately, there is a limit to these arguments.  The Galaxy becomes completely opaque to photons with energy of $10^{24} \eV$ and higher, because single photons can pair produce off the Galactic magnetic field (\citealt{Stecker03}; see also \S~3 of \citealt{Erber66}).  Therefore, we can expect no pair-production bounds on the extragalactic radio background below about 250 Hz.  Any such bounds also assume Lorentz invariance holds at these high energies; some theories do not predict this \citep[see][]{AmelinoCamelia01,Galaverni08}.

\section{Conclusion}
\label{sec:Conclusion}
\begin{figure*}
\centerline{\includegraphics[width=18cm]{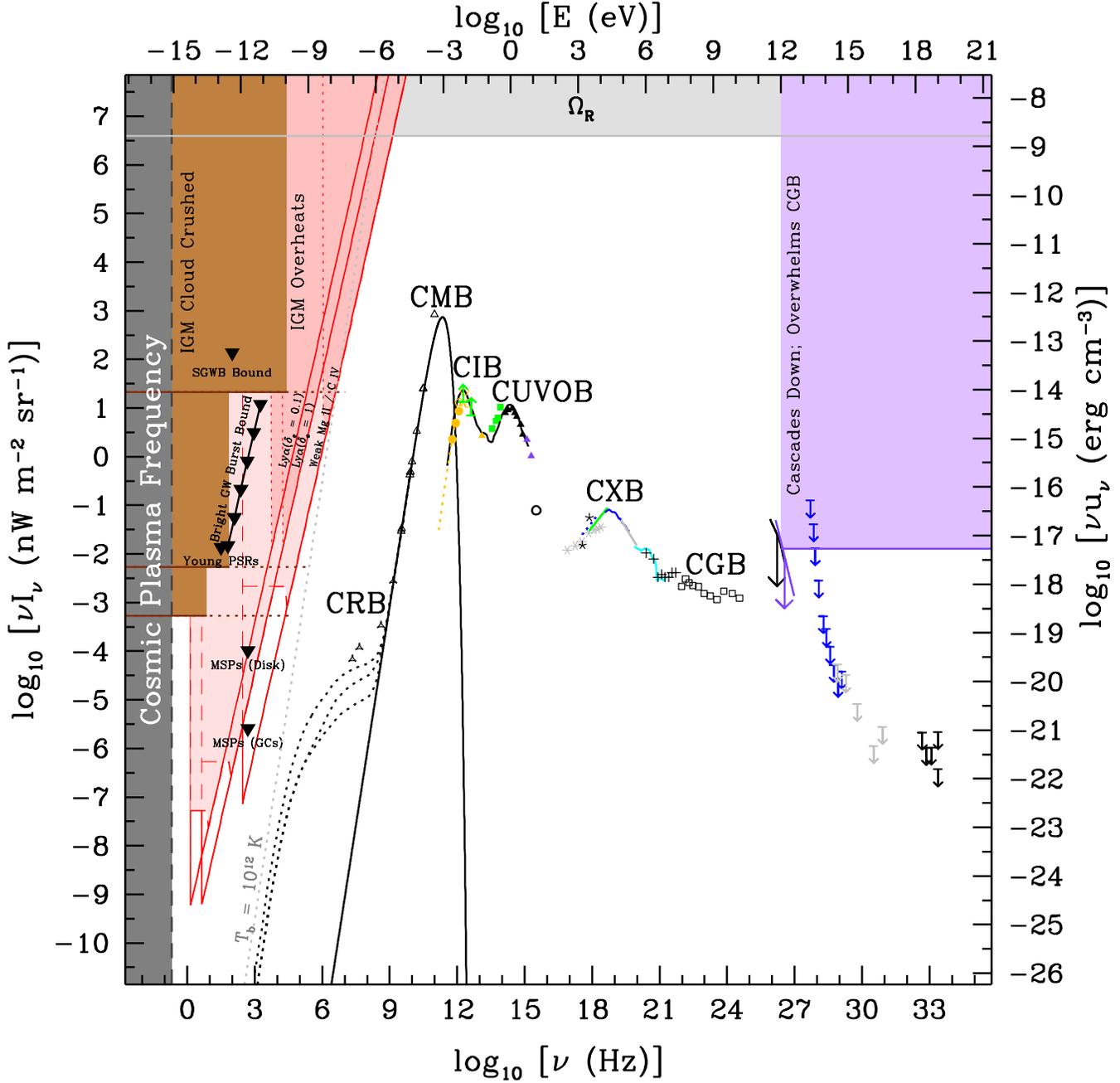}}
\caption{The $z = 0$ cosmic backgrounds for the electromagnetic spectrum.  The cosmic plasma frequency (assuming $\delta_e \approx 0.002$) makes propagation impossible at the lowest frequencies (\emph{dark grey}); we expect there to be \emph{no} radio background below this cutoff.  The $\Omega_R$ bound \citep{Zentner02} is in grey at top (\S~\ref{sec:OmegaR}).  Free-free absorption bounds (\S~\ref{sec:FFAbsorption}) from Ly$\alpha$ clouds of $\delta = 0.1$ and $1.0$ and from a weak Mg II/C IV absorber are shown in pink/red, assuming $J_{\nu}$ is a bump at $\nu$.  The solid red bounds are for the case when the radiation fills the cloud evenly.  The shaded regions are bounds on an incident external radiation field; lighter shading (\emph{dashed boundaries}) for no scattering in the cloud, while darker shading (\emph{dotted boundaries}) when eq.~\ref{eqn:mfpScattering} describe the scattering.
Radiation pressure can crush a cloud (\S~\ref{sec:CloudCrush}); the bounds on cloud crushing are shown in brown.  The solid line and shading assumes there is no scattering, while the dotted line assumes that eq.~\ref{eqn:mfpScattering} describe the scattering.  The radiation pressure bounds are bolometric below $\nu_{\rm int}$.  
I also plot naive upper expectations on the radio background expected from several sources (\S~\ref{sec:SubMHzSources}), not accounting for IGM absorption: the maximum synchrotron brightness temperature ($T_b \approx 10^{12}\ \Kelv$), pulsars, and gravitational wave conversion.  
See Table~\ref{table:BigFigKey} for a full legend with references.
\label{fig:EMBackgrounds}}
\end{figure*}

The extragalactic sub-MHz background is invisible to direct observation from Earth, but we can still detect its effects on intergalactic matter.  I have placed new limits on the magnitude of the sub-MHz radio background, using various IGM phases and clusters as radio detectors.  Figure~\ref{fig:EMBackgrounds} (full legend in Table~\ref{table:BigFigKey}) summarises the bounds on the radio background from the IGM thermal state (\S~\ref{sec:FFUBound}; \emph{red}) and the radiation pressure exerted on the IGM  (\S~\ref{sec:CloudCrush}; \emph{brown}).  A sub-MHz background with an energy density as large as the CMB at any frequency is easily ruled out, and energy densities comparable to the cosmic starlight backgrounds are also not allowed at almost all frequencies.

\begin{table*}
\begin{minipage}{140mm}
\caption{Legend for Figure~\ref{fig:EMBackgrounds}.}
\begin{tabular}{llll}
\hline
Wavelength band & Method/Instrument & Reference & Symbol\\
\hline
All & $\Omega_R$ & \citet{Zentner02} & \emph{light grey shading}\\
Sub-MHz radio & IGM pressure & This work (\S~\ref{sec:CloudCrush}) & \emph{brown shading}\\
      & IGM thermal state & This work (\S~\ref{sec:FFUBound}) & \emph{pink shading}\\
Radio & Theoretical prediction & \citet{Protheroe96} & \emph{dotted black lines}\\
      & ARCADE2 & \citet{Fixsen09} & \emph{open triangles}\\
IR to UV & Theoretical prediction & \citet*{Franceshini08} & \emph{solid black line} \\
Infrared & FIRAS & \citet{Fixsen98} & \emph{orange dotted line}\\
      & BLAST & \citet{Marsden09} & \emph{orange circles}\\
      & DIRBE & \citet{Wright04} & \emph{orange pentagons}\\
      & \emph{Spitzer} & \citet{Dole06} & \emph{green lower limits}\\
      & \emph{Spitzer} & \citet{Papovich04} & \emph{orange triangle}\\
      & \emph{Spitzer} & \citet{Savage05} & \emph{green squares}\\
Optical & Galaxy counts with \emph{Hubble} & \citet{Madau00} & \emph{filled triangles} \\
UV & GALEX & \citet{Xu05} & \emph{violet triangles}\\
EUV & Ly$\alpha$ forest ionization & \citet{Shull99} & \emph{open circle}\\
X-rays & \emph{XMM-Newton} (Lockman hole) & \citet{Worsley05} & \emph{grey 6-stars}\\
      & \emph{Chandra} & \citet{Hickox06} & \emph{5-stars}\\
      & \emph{Swift} & \citet{Moretti09} & \emph{dotted blue line}\\
      & \emph{Swift} & \citet{Ajello08} & \emph{solid blue line}\\
      & RXTE & \citet{Revnivtsev03} & \emph{solid green line}\\
      & HEAO1 & \citet{Kinzer97} & \emph{grey solid line}\\
MeV $\gamma$-rays & SMM & \citet{Watanabe00} & \emph{solid cyan line}\\
      & COMPTEL & \citet{Weidenspointner00} & \emph{crosses}\\
GeV $\gamma$-rays & EGRET & \citet*{Strong04} & \emph{open squares}\\
$\ge \TeV$ $\gamma$-rays & GeV background & \citet{Coppi97} & \emph{violet shading}\\
TeV $\gamma$-rays & HESS & \citet{HESS09} & \emph{solid black line}\\
      & HESS & \citet{Aharonian08} & \emph{solid violet line}\\
      & GRAPES-3 & \citet{Hayashi03} & \emph{blue arrows}\\
PeV $\gamma$-rays & CASA-MIA & \citet{Chantell97} & \emph{grey arrows}\\
EeV $\gamma$-rays & Auger & \citet{Auger09} & \emph{black arrows}\\
      & Auger & \citet{Abraham08} & \emph{black arrows}\\
\hline
\end{tabular}
\label{table:BigFigKey}
\\The PeV $\gamma$-ray limits do not correct for pair-production absorption from the CMB.\\
\end{minipage}
\end{table*}

Low frequency radio waves can heat the IGM through free-free absorption.  Observations of the IGM thermal state constrain the amount of heating from the extragalactic sub-MHz radio background.  Free-free absorption bounds (\S~\ref{sec:FFUBound}) are potentially the strongest of all of the limits, but are highly model dependent at low frequencies.  If we are considering a radio bath that pervades the entire cloud evenly (\emph{dashed red lines}), then the entire cloud is heated up and the bounds are very strong at low frequency in terms of energy density.  In fact, the energy density constraints within the Lyman-$\alpha$ forest just above its plasma frequency would be the strongest of any photon energy, as seen in Figure~\ref{fig:EMBackgrounds}.  If a background was simply incident on the IGM clouds, then when the cloud becomes optically thick, the outside of the cloud will be heated but the interior will not.  In this case, if there is no scattering of radio waves within the cloud (\emph{solid red lines, light pink shading}), the clouds usually remain optically thin down to sub-kHz frequency, and the energy density bounds remain strong.  However, scattering will increase the effective absorption optical depth.  Naively applying the scattering mean free path in equation~\ref{eqn:mfpScattering} considerably weakens the energy density bounds (\emph{dotted red lines, darker pink shading}).  However, the approximations in equation~\ref{eqn:mfpScattering} may break down at high scattering optical depth (\ref{sec:IGMPropagation}; \citealt{Cohen74}).  I have also not considered the opacity of any remaining shell of `evaporated' material around an optically thick cloud.  In order to set firm limits from free-free absorption, we need to understand the radiative transfer of sub-MHz radio waves through the IGM better.

I have also set an upper limit on the luminosity density \emph{within} each IGM phase from free-free absorption (\S~\ref{sec:FFEpsilonBound}), assuming the radio waves fill the IGM cloud evenly.  From the existence and temperature of the Lyman-$\alpha$ forest, I infer that at frequencies of $5 - 100$ Hz, these clouds have a maximum emissivity of $6 \times 10^4\ \Lsun\ \Mpc^{-3}$.  Again, a better understanding of the radiative transfer and the scattering in particular of sub-MHz radio waves is needed to set more firm limits on the luminosity density of the Universe at low frequencies.

At the lowest frequencies, there is a window in the free-free absorption constraints.  This is because voids are extremely underdense, with a very low plasma frequency.  Sub-Hz radio could be generated in the voids and would simply reflect off the more condensed structures that would otherwise be heated by them.  

Free-free absorption constraints are weak because scattering or absorption of the radio waves shields the interior of each IGM cloud.  At the very lowest frequencies, below the plasma frequency of the cloud, the radio waves simply reflect off it.  There are none the less constraints even at these lowest frequencies, because the reflection, scattering, or absorption of these waves squeezes IGM clouds (\S~\ref{sec:CloudCrush}).  These radiation pressure bounds (Figure~\ref{fig:EMBackgrounds}, \emph{brown}) are somewhat weak but are still strong enough to rule out an average sub-Hz background as large as the starlight backgrounds.  Unlike the free-free absorption bounds, the maximum $u_{\nu}$ are not model dependent.  However, the frequency range over which they are applicable also depends on the radio scattering properties of the IGM.  At the very least, the radiation pressure bounds apply until the cloud is optically thin to free-free absorption.

Galaxy clusters have hot and dense gas, which makes them poor free-free absorption detectors for any sub-MHz radio waves within them.  However, they also contain magnetic fields and cosmic rays, which can interact with sub-MHz radio waves in additional ways and provide additional limits (Figure~\ref{fig:ClusterLimits}).  Low frequency radio waves above the Razin frequency ($\sim 20\ \kHz$) can actually heat CR electrons/positrons, and would create a peak into the observed MHz to GHz synchrotron radio spectra of clusters (\S~\ref{sec:SynchHeat}).  The lack of such a peak rules out sub-MHz radio backgrounds as small as $\sim 10^{-15} \ergcm3$ in the Coma cluster, comparable to the energy density in starlight.  CR electrons/positrons can also Inverse Compton scatter low frequency radio waves to observable MHz to GHz frequencies (\S~\ref{sec:ICBound}).  The observed radio spectrum again constrains the sub-MHz radio background in Coma to be as small as $\sim 10^{-15} \ergcm3$ for $\nu \la \kHz$.  

Some relatively weak statements can be made about whether the backgrounds described in \S~\ref{sec:SubMHzSources} exist.  The most exotic sources of sub-MHz radio waves are constrained.  The radiation pressure bound from weak Mg II/C IV absorbers are sufficiently strong to exclude a radio background as large as the LIGO upper limits on a 100 Hz stochastic gravitational wave background.  A background of pulsar waves, if they somehow escaped into the IGM and did not suffer absorption, would be weaker still; the free-free absorption bounds without scattering are strong enough to rule out such backgrounds from all young pulsars and all MSPs in Galactic discs.  These maximum estimates of the sub-MHz radio background are probably unrealistic anyway (see the discussion in \S~\ref{sec:ExoticSources}).  Unfortunately, a more realistic synchrotron background also seems to be out of reach by the free-free absorption bounds; a $T_b \le 10^{12} \Kelv$ background is ruled out only for $\nu \ga \MHz$, which can already be directly observed.  However, if the scattering properties of the IGM are similar to those in the ISM, such that eq.~\ref{eqn:mfpScattering} holds, then the radiation pressure bounds from the Lyman-$\alpha$ forest will rule out maximal synchrotron backgrounds at $\sim 100\ \kHz$.

There are several ways to reduce the uncertainties in these bounds.  Knowledge of the low frequency scattering properties of the IGM is essential for the free-free absorption bounds, which would otherwise be strong (\emph{solid red lines} in Figure~\ref{fig:EMBackgrounds}).  This knowledge can also help us determine the frequency range the radiation pressure bounds apply over.  Strong scattering will \emph{weaken} the free-free absorption bounds, because radiation cannot diffuse deep into the cloud and heat it; but strong scattering \emph{strengthens} the radiation pressure bounds, because radiation can then efficiently couple with the cloud exterior and squeeze it.  Low frequency radio observations of galaxy clusters and other environments with CRs are especially useful for the Inverse Compton and synchrotron heating bounds.  Observations of extragalactic ultra high energy photons, if they exist, would set extremely strong constraints on the extragalactic sub-MHz background (\S~\ref{sec:UHELimits}), especially at high frequencies where the other bounds are weakest.

It would be a simple matter to apply similar constraints to sub-MHz emission within the Galaxy itself, which will be done in a future paper.  Not only are the thermal properties of the interstellar medium relatively well characterised, but the Galaxy has a well known CR electron spectrum and magnetic field.  Therefore, we could apply synchrotron heating and IC upscattering arguments, which are not dependent on the scattering of low frequency radio waves.  This would allow us to probe distant regions of the Galaxy that are not visible at low radio frequencies because of free-free absorption, such as the Galactic Centre, which is obscured by free-free absorption at frequencies as high as 330 MHz \citep[e.g.,][]{Pedlar89}.  

Although the bounds in this paper may not strongly constrain expected sources like a synchrotron background, the extragalactic sub-MHz sky is not \emph{completely} unknowable.  Instead of disregarding the low frequency emission of radio sources, it is possible to consider the effects of the emission on their surroundings.  These effects may prove to be important to our understanding of the regions around sub-MHz sources.  Even if the extragalactic sub-MHz sky is forever invisible to us directly, its presence can still be seen.

\section*{Acknowledgments}
I thank Chris Kochanek for a critical reading of this paper.  I would also like to thank Todd Thompson and John Beacom for encouragement, readings, and discussions.  This project was supported in part by an Alfred P. Sloan Fellowship to Todd Thompson and NSF CAREER Grant PHY-0547102 through John Beacom.

\end{document}